\documentclass[prd,preprint,tightenlines,floatfix,showpacs,preprintnumbers,nofootinbib,eqsecnum]{revtex4}
 \usepackage[dvips,final]{graphicx}
  \usepackage{amssymb}
   \usepackage{amsmath}
    \usepackage{amsfonts}
     \usepackage{epsfig}
      \usepackage{bm}

\def\Im{\,{\rm Im}\,}
\def\mb{\,\mbox{mb}}
\def\fm{\,\mbox{fm}}
\def\GeV{\,\mbox{GeV}}
\def\TeV{\,\mbox{TeV}}
\def\Pom{{ I\!\!P}}
 \def\Reg{{ I\!\!R}}
\newcommand\la{\langle}
 \newcommand\ra{\rangle}
 \newcommand\beq{\begin{equation}}
 
 \newcommand\eeq{\end{equation}}
 \newcommand\beqn{\begin{eqnarray}}
 \newcommand\eeqn{\end{eqnarray}}

\begin{document}

\begin{flushright}
LU TP 12-14\\
USM-TH-300
\end{flushright}

\title{Diffractive gauge bosons production beyond QCD factorisation}

\author{R. S. Pasechnik}
 \email{Roman.Pasechnik@thep.lu.se}
 \affiliation{
Department of Astronomy and Theoretical Physics, Lund University, SE
223-62 Lund, Sweden}

\author{B. Z. Kopeliovich, and I. K. Potashnikova}
 \affiliation{ Departamento de F\'{\i}sica Universidad T\'ecnica
 Federico Santa Mar\'{\i}a; and\\
 Instituto de Estudios Avanzados en Ciencias e Ingenier\'{\i}a; and\\
 Centro Cient\'ifico-Tecnol\'ogico de Valpara\'iso;\\
 Casilla 110-V, Valpara\'iso, Chile}

\begin{abstract}
We discuss single diffractive gauge bosons
($\gamma^*,\,W^{\pm},\,Z$) production in proton-proton collisions at
different (LHC and RHIC) energies within the color dipole approach.
The calculations are performed for gauge bosons produced at forward
rapidities. The diffractive cross section is predicted as function
of fractional momentum and invariant mass of the lepton pair. We
found a dramatic breakdown of the diffractive QCD factorisation
caused by an interplay of hard and soft interactions. Data from the
CDF experiment on diffractive production of $W$ and $Z$ are well
explained in a parameter free way.
\end{abstract}


\maketitle

\section{Introduction}

The characteristic feature of diffractive processes at high energies
is presence of a large rapidity gap between the remnants of the beam
and target. The general theoretical framework for such processes was
formulated in the pioneering works of  Glauber \cite{Glauber},
Feinberg and Pomeranchuk \cite{FP56}, Good and Walker \cite{GW}. The
new {\it diffractive} state is produced only if different Fock
components of the incoming plane wave, which are the eigenstates of
interaction, interact differently with the target. As a results in a
new combination of the Fock components, which can be
projected to a new physical state (e.g. see review on QCD
diffraction Ref.~\cite{KPSdiff}).

The main difficulty in the formulation of a theoretical QCD-based
framework for diffractive scattering arises from the fact that  it
is essentially contaminated by soft, non-perturbative interactions.
For example, diffractive deep-inelastic scattering (DIS),
$\gamma^*p\to Xp$, although it is a higher twist process, is
dominated by soft interactions \cite{povh}. Within the dipole
approach \cite{zkl} such a process looks like elastic scattering of
$\bar qq$ dipoles of different sizes, and of higher  Fock states
containing more partons. Although formally the process $\gamma^*\to
X$ is an off-diagonal diffraction, it does not vanish in the limit
of unitarity saturation, the so called black disc limit. This
happens because the photon distribution functions and hadronic wave
functions are not orthogonal. Such a principal difference between
diffractive processes in DIS and hadronic collisions is one of the
reasons for breakdown of diffractive QCD factorization based e.g. on
the Ingelman-Schlein model \cite{ISh}. In particular the cross
section of diffractive production of the $W$ boson was found in the
CDF experiment \cite{cdf1,CDF-WZ} to be six times smaller than was
predicted relying on factorization and HERA data \cite{dino}. The
phenomenological models based on assumptions of the diffractive
factorisation, which are widely discussed in the literature (see
e.g. Refs.~\cite{Szczurek,Beatriz}), predict a significant increase
of the ratio of the diffractive to inclusive gauge bosons production
cross sections with energy. This is supposed to be tested soon at
the LHC.

The process under discussion, diffractive abelian radiation of
electroweak gauge bosons, is the real off-diagonal diffraction. It
vanishes in the black-disc limit, and may be strongly suppressed by
the absorptive corrections even being far from the unitarity bound.
The suppression caused by the absorptive corrections, also known as
the survival probability of a large rapidity gap, is related to the
initial and final state interactions. Usually the survival
probability is introduced in the diffractive cross section in a
probabilistic way and is estimated
in oversimplified models, like eikonal, quasi-eikonal,
two-channel approximations, etc. The advantage of the dipole
approach is the possibility to calculate directly (although in a
model-dependent way) the full diffractive amplitude, which contains
all the absorption corrections, because it employs the phenomenological
dipole cross section fitted to data. Below we explicitly single out
from the diffractive amplitude the survival probability amplitude as
a factor.

Another source of factorization breaking is the simple observation
that diffractive abelian radiation by a quark vanishes in the
forward direction (zero momentum transfer to the target)
\cite{KST-par}. Indeed, the Fock components of the quark with or
without the abelian boson ($\gamma^*,\ Z,\ W$, Higgs boson) interact
with the same total cross sections, because only the quark interacts
strongly. Therefore, after integration of the amplitude over impact
parameter the Fock state decomposition of the projectile remains
unchanged, and only elastic $qp$ scattering is possible. Notice that
nonabelian radiation (gluons) does not expose this property, because
the Fock components $|q\rangle$ and $|q\,g\rangle$, although have
the same color, interact differently \cite{KST-par,KPST07}.

In the case of $pp$ collisions the directions of propagation of the
proton and its quarks do not coincide. Already this is sufficient to
get a nonvanishing diffractive abelian radiation in forward
scattering. Moreover, interaction with the spectator partons opens
new possibilities for diffractive scattering, namely the color
exchange in interaction of one projectile parton, can be compensated
(neutralized) by interaction of another projectile parton. It was
found in Refs.~\cite{KPST06,our-DDY} that this contribution leads to
a dominant contribution to the diffractive abelian radiation in the
forward direction. This mechanism, leading to a dramatic violation
of diffractive QCD factorisation, is under consideration in the
present paper. The breakdown of the diffractive (Ingelman-Schlein)
QCD factorisation is a result of the interplay between the soft and
hard interactions, which considerably affects the corresponding
observables \cite{KPST06}. Recently, such an effect has been
analyzed in the diffractive Drell-Yan process \cite{KPST06,our-DDY},
and here we extend our study of this interesting phenomenon to a
more general case -- the diffractive gauge boson production.

\section{Diffractive gauge bosons production amplitude}

Consider first the general formalism for diffractive radiation of the electroweak gauge bosons,
$\gamma^*,\,Z^0,\,W^{\pm}$, within the color dipole
approach. Let us start with consideration of the distribution functions
for the Fock states contributing to heavy gauge bosons radiation
by a quark (valence or sea) in the projectile proton.

\subsection{Gauge bosons radiation by a quark}

The $q_f\to q_f \gamma^*$ transition amplitude is given by the
vector $\gamma^* q$ coupling only, i.e.
\begin{eqnarray}\label{Agam}
&&T(q_f\to q_f \gamma^*)=-ie Z_q\varepsilon^{\mu}_{\lambda_{\gamma^*}}
\bar{u}_f\gamma_{\mu}u_f\,,
\end{eqnarray}
where $Z_f$ is the quark charge, ${\bar u}_f$ and $u_f$ are spinors
for the quark of the flavour $f$ in the final and initial states,
respectively.

The couplings of $Z^0$ and $W^{\pm}$ bosons to quarks contain both
vector and axial-vector parts. The $q_f\to q_f Z^0$ transition
amplitude is given by
\begin{eqnarray}\label{AZ}
&&T(q_f\to q_f
Z^0)=\frac{-ie}{\sin\,2\theta_W}\,\varepsilon^{\mu}_{\lambda_Z}
\bar{u}_f[g_{v,f}^Z\gamma_{\mu}-g_{a,f}^Z\gamma_{\mu}\gamma_5]u_f,
\end{eqnarray}
while the $q_f\to q_{f'} W^{\pm}$ amplitudes read,
\begin{eqnarray}\label{AW}
&&T(q_{f_u}\to q_{f_d}
W^+)=\frac{-ie}{2\sqrt{2}\sin\theta_W}\,V_{f_uf_d}\,\varepsilon^{\mu}_{\lambda_W}
\bar{u}_{f_u}[g_{v,f}^W\gamma_{\mu}-g_{a,f}^W\gamma_{\mu}\gamma_5]u_{f_d},\\
&&T(q_{f_d}\to q_{f_u}
W^-)=\frac{-ie}{2\sqrt{2}\sin\theta_W}\,V_{f_df_u}\,\varepsilon^{\mu}_{\lambda_W}
\bar{u}_{f_d}[g_{v,f}^W\gamma_{\mu}-g_{a,f}^W\gamma_{\mu}\gamma_5]u_{f_u},
\nonumber
\end{eqnarray}
for the {\it up} ($f_u=u,c,t$) and {\it down} ($f_d=d,s,b$) quarks, respectively.
Here, $V_{f_uf_d}$ is the CKM matrix element corresponding to
$f_u\to f_d$ transition, and $\theta_{W}$ is the Weinberg angle. The
weak mixing parameter, $\sin^2{\theta_W}$, is related at the tree level
to $G_F,\,M_Z$ and $\alpha_{em}$ by
$\sin^2{\theta_W}=4\pi\alpha_{em}/\sqrt{2}G_F M_Z^2$ (we adopt here
$\alpha_{em}(m_Z)\simeq 1/127.934$). The vector couplings at the tree
level are
\begin{eqnarray}\label{vec}
g_{v,f_u}^Z=\frac12-\frac43\sin^2\theta_W,\qquad
g_{v,f_d}^Z=-\frac12+\frac23\sin^2\theta_W,\qquad g_{v,f}^W=1\,;
\end{eqnarray}
whereas axial-vector couplings are
\begin{eqnarray}\label{axial}
g_{a,f_u}^Z=\frac12,\qquad g_{a,f_d}^Z=-\frac12\,,\qquad
g_{a,f}^W=1\,.
\end{eqnarray}

Heavy gauge boson polarization vectors describing transverse (T),
$\lambda_G=\pm1$, and longitudinal (L), $\lambda_G=0$, polarization
states are defined in light-cone coordinates\footnote{As usual,
the light-cone 4-vector $p$ is defined as
$p=(p^+,\,p^-,\,\vec{p})$, where $p^{\pm}=p^0\pm p^z$.} as
\begin{eqnarray}\label{pols}
&&\varepsilon_{\lambda=\pm}=(0,\,0,\,\vec{\varepsilon}_{\lambda=\pm}),\qquad
\vec{\varepsilon}_{\lambda=\pm}=\mp\frac{1}{\sqrt{2}}(1,\,\pm i),\\
&&\varepsilon_{\lambda=0}=\left(\frac{q^+}{M},\,-\frac{M}{q^+},\,\vec{0}\right).
\end{eqnarray}
Note that we work in the physical (unitary) gauge. The calculations
are performed in the high energy limit, i.e. in the limit where
$q^+$ is much larger than all other scales.

Let us start with the radiation of a heavy gauge boson by a quark
interacting with a proton target. We assume that the longitudinal
momentum of the projectile is not changed significantly by the soft interaction at
high energies. In the high energy limit the corresponding $s$ and
$u$-channel amplitudes of the gauge boson bremsstrahlung in the
quark-target scattering can be written as follows (cf. diffractive
DY amplitude in Ref.~\cite{our-DDY}),
\begin{eqnarray} \nonumber
&&{\cal M}_{s}\simeq -i\sqrt{4\pi} {\cal
C}_q^G\,\alpha(1-\alpha)\,\varepsilon^{\mu}_{\lambda}\sum_{\sigma}\frac{{\bar
u}_{\sigma_2}(p_2)[g_{v,f}^G\gamma_{\mu}-g_{a,f}^G\gamma_{\mu}\gamma_5]u_{\sigma}(p_2+q)}{\alpha^2l_{\perp}^2+\eta^2}\,{\cal
A}_{\sigma\sigma_1}(k_{\perp}),\\&&{\cal M}_{u}\simeq i\sqrt{4\pi}
{\cal
C}_q^G\,\alpha\,\varepsilon^{\mu}_{\lambda}\sum_{\sigma}\frac{{\bar
u}_{\sigma}(p_1-q)[g_{v,f}^G\gamma_{\mu}-g_{a,f}^G\gamma_{\mu}\gamma_5]u_{\sigma_1}(p_1)}
{\alpha^2(\vec{l}_{\perp}+\vec{k}_{\perp})^2+\eta^2}\,{\cal
A}_{\sigma\sigma_2}(k_{\perp}),\label{us}
\end{eqnarray}
where $G=\gamma,\,W^{\pm},\,Z$ is the gauge boson under
consideration; $\eta^2=(1-\alpha)M^2+\alpha^2m_q^2$; $\alpha$ is the
fractional light-cone momentum carried by the gauge boson, which has
invariant mass $M$. The vector ($v$) and axial-vector ($a$)
couplings $g^{Z,W}_{v/a,f}$ are defined in Eqs.~(\ref{vec}) and
(\ref{axial}), whereas our notations imply $g^{\gamma}_{v,f}=1$ and
$g^{\gamma}_{a,f}=0$; $\sigma_{1,2}$ are the helicities of initial
and final quarks, respectively;
$\vec{k}_{\perp}=\vec{p}_{2\perp}-\vec{p}_{1\perp}+\vec{q}_{\perp}$
is the transverse momentum of the exchanged gluon; and
$\vec{l}_{\perp}=\vec{p}_{2\perp}-(1-\alpha)\vec{q}_{\perp}/\alpha$
is the transverse momentum of the final quark in the frame, where
$z$-axis is parallel to the gauge boson momentum. The amplitude
${\cal A}$ for scattering of the quark on the nucleon in the target
rest frame has the following approximate form, \cite{BHQ}
\begin{eqnarray*}
{\cal A}_{\sigma\sigma_1}(\vec{k}_{\perp})\simeq
2p_1^0\delta_{\sigma\sigma_1}{\cal V}_q(\vec{k}_{\perp}),\qquad
{\cal A}_{\sigma\sigma_2}(\vec{k}_{\perp})\simeq
2p_2^0\delta_{\sigma\sigma_2}{\cal V}_q(\vec{k}_{\perp})\,,
\end{eqnarray*}
where the factorized universal amplitude ${\cal
V}_q(\vec{k}_{\perp})$ does not depend on energy and helicity state
of the quark. The coupling factor ${\cal C}_f^G,\,f=q,l$, introduced
in Eq.~(\ref{us}), is defined for $G=\gamma^*$, $Z^0$ and $W^{\pm}$
bosons, respectively, as
\begin{eqnarray} \nonumber
{\cal C}^{\gamma}_f=\sqrt{\alpha_{em}} Z_f,\qquad {\cal
C}^Z_f=\frac{\sqrt{\alpha_{em}}}{\sin 2\theta_W},\qquad {\cal
C}^{W^+}_f=\frac{\sqrt{\alpha_{em}}}{2\sqrt{2}\sin\theta_W}V_{f_uf_d},\qquad
{\cal
C}^{W^-}_f=\frac{\sqrt{\alpha_{em}}}{2\sqrt{2}\sin\theta_W}V_{f_df_u}\,,
\end{eqnarray}
where $\alpha_{em}=e^2/(4\pi)=1/137$ is the electromagnetic coupling
constant. In the case of gauge boson couplings to leptons we should
substitute $V_{f_uf_d}=V_{f_df_u}=1$.

Eventually, we can switch to impact parameter space performing the Fourier
transformation over $\vec{l}_{\perp}$ and $\vec{k}_{\perp}$, and
write down the total amplitude  $M_q$ for gauge boson radiation in
quark-proton scattering as follows,
\begin{eqnarray*}
&&M_q^{\mu}(\vec{b},\vec{r})=-2ip_1^0\,\sqrt{4\pi}\,\frac{\sqrt{1-\alpha}}{\alpha^2}\,
\Psi^{\mu}_{V-A}(\vec{r},\alpha,M)\cdot
\Big[V_q(\vec{b})-V_q(\vec{b}+\alpha\vec{r})\Big],\\
&&V_q(\vec{b})=\int\frac{d^2k_{\perp}}{(2\pi)^2}e^{-i\vec{k}_{\perp}\cdot
\vec{b}}{\cal V}_q(\vec{k}_{\perp}),\quad
\Psi^{Z,W}_{V-A}(\vec{r},\alpha,M)=\Psi^{Z,W}_V(\vec{r},\alpha,M)-\Psi^{Z,W}_A(\vec{r},\alpha,M)\,,
\end{eqnarray*}
where $\alpha\vec{r}$ is the transverse separation
between the initial and final quarks;
$\Psi^{\mu}_{V/A}(\vec{r},\alpha,M)$ are the light-cone distribution
functions of the vector $q\to Vq$ and axial-vector $q\to Aq$
transitions in the mixed representation defined as,
\begin{eqnarray*}
&&\Psi^{\mu}_V(\vec{r},\alpha,M)={\cal C}^G_q g_{v,q}^G\,
\alpha^3\sqrt{1-\alpha}\int\frac{d^2l_{\perp}}{(2\pi)^2}e^{-i\vec{l}_{\perp}\cdot
\alpha\vec{r}}\,\frac{{\bar
u}_{\sigma_2}(p_f)\gamma^{\mu}u_{\sigma}(p_2+q)}{\alpha^2l_{\perp}^2+\eta^2},\\
&&\Psi^{\mu}_A(\vec{r},\alpha,M)={\cal C}^G_q g_{a,q}^G\,
\alpha^3\sqrt{1-\alpha}\int\frac{d^2l_{\perp}}{(2\pi)^2}e^{-i\vec{l}_{\perp}\cdot
\alpha\vec{r}}\,\frac{{\bar
u}_{\sigma_2}(p_2)\gamma^{\mu}\gamma_5u_{\sigma}(p_2+q)}{\alpha^2l_{\perp}^2+\eta^2}\,.
\end{eqnarray*}
For an unpolarized initial quark the interference terms between the
vector and axial-vector wave functions cancel each other, i.e.
\begin{eqnarray}
\sum_{\sigma_1,\sigma_2}\Psi^{\lambda}_{V-A}(\alpha,\vec{\rho}_1)
\Psi^{\lambda*}_{V-A}(\alpha,\vec{\rho}_2)=\Psi^{\lambda}_V(\alpha,\vec{\rho}_1)
\Psi^{\lambda*}_V(\alpha,\vec{\rho}_2)+\Psi^{\lambda}_A(\alpha,\vec{\rho}_1)
\Psi^{\lambda*}_A(\alpha,\vec{\rho}_2)\,.
\label{V-A}
\end{eqnarray}
The bilinear combinations of the vector $V$ and
axial-vector $A$ light-cone distribution functions, corresponding to
radiation of longitudinally ($\lambda=0$) and transversely
($\lambda=\pm1$) polarized gauge bosons have the form\footnote{In the case of
a heavy photon $\gamma^*$ bremsstrahlung by a quark such formulae were derived in
Refs.~\cite{deriv1,BHQ,deriv2}.},
\begin{eqnarray}\label{VV}
&&\Psi^{T}_{V}(\alpha,\vec{\rho}_1)
\Psi^{T*}_{V}(\alpha,\vec{\rho}_2)= \sum_{\lambda=\pm
1}\frac{1}{2}\sum_{\sigma_1,\sigma_2}
\epsilon^*_\mu(\lambda)\Psi^{\mu}_{V}(\alpha,\vec{\rho}_1)
\epsilon_\nu(\lambda)\Psi^{\nu*}_{V}(\alpha,\vec{\rho}_2)\\
\nonumber &&\qquad=\frac{{\cal C}_q^2(g^{G}_{v,q})^2}{2\pi^2}\Bigg\{
     m_q^2 \alpha^4 {\rm K}_0\left(\eta \rho_1\right)
     {\rm K}_0\left(\eta \rho_2\right)+ \left[1+\left(1-\alpha\right)^2\right]\eta^2
   \frac{\vec{\rho}_1\cdot\vec{\rho}_2}{\rho_1\rho_2}
     {\rm K}_1\left(\eta \rho_1\right)
     {\rm K}_1\left(\eta \rho_2\right)\Bigg\},\\ \nonumber
&&\Psi^{L}_{V}(\alpha,\vec{\rho}_1)
\Psi^{L*}_{V}(\alpha,\vec{\rho}_2)=
\frac{1}{2}\sum_{\sigma_1,\sigma_2}
\epsilon^*_\mu(\lambda=0)\Psi^{\mu}_{V}(\alpha,\vec{\rho}_1)
\epsilon_\nu(\lambda=0)\Psi^{\nu*}_{V}(\alpha,\vec{\rho}_2)\\
&&\qquad=\frac{{\cal C}_q^2(g_{v,q}^G)^2}{\pi^2}M^2
\left(1-\alpha\right)^2
  {\rm K}_0\left(\eta \rho_1\right)
     {\rm K}_0\left(\eta \rho_2\right)\,. \nonumber
\end{eqnarray}
\begin{eqnarray} \label{AA}
&&\Psi^{T}_{A}(\alpha,\vec{\rho}_1)\Psi^{T*}_{A}(\alpha,\vec{\rho}_2)=
 \\ \nonumber &&\qquad=\frac{{\cal
   C}_q^2(g_{a,q}^G)^2}{2\pi^2}\Bigg\{
     m_q^2 \alpha^2(2-\alpha)^2 {\rm K}_0\left(\eta \rho_1\right)
     {\rm K}_0\left(\eta \rho_2\right)+ \left[1+\left(1-\alpha\right)^2\right]\eta^2
   \frac{\vec{\rho}_1\cdot\vec{\rho}_2}{\rho_1\rho_2}
     {\rm K}_1\left(\eta \rho_1\right)
     {\rm K}_1\left(\eta \rho_2\right)\Bigg\},\\ \nonumber
&&\Psi^{L}_{A}(\alpha,\vec{\rho}_1)
\Psi^{L*}_{A}(\alpha,\vec{\rho}_2)=\frac{{\cal
C}_q^2(g_{a,q}^G)^2}{\pi^2}\frac{\eta^2}{M^2}\Bigg\{\eta^2
  {\rm K}_0\left(\eta \rho_1\right)
     {\rm K}_0\left(\eta \rho_2\right)+\alpha^2m_q^2\frac{\vec{\rho}_1\cdot\vec{\rho}_2}{\rho_1\rho_2}
     {\rm K}_1\left(\eta \rho_1\right)
     {\rm K}_1\left(\eta \rho_2\right)\Bigg\}.
\end{eqnarray}
where the averaging over helicity of the initial quark is performed.

\subsection{Forward diffractive radiation from a dipole}

The amplitude of diffractive gauge boson radiation by a
quark-antiquark dipole does not vanish in forward direction, unlike
the radiation by a single quark \cite{KST-par,KPST06}. This can be
understood as follows. According to the general theory of
diffraction \cite{Glauber,FP56,GW,KPSdiff}, the off-diagonal
diffractive channels are possible only if different Fock components
of the projectile (eigenstates of interaction) interact with
different elastic amplitudes. Clearly, the two Fock states
consisting of just a quark and of a quark plus a gauge boson
interact equally, if their elastic amplitudes are integrated over
impact parameter. Indeed, when a quark fluctuates into a state
$|qG\rangle$ containing the gauge boson $G$, with the transverse
quark-boson separation $\vec{r}$, the quark gets a transverse shift
$\Delta\vec{r}=\alpha\vec{r}$. The impact parameter integration
gives the forward amplitude. Both Fock states $|q\rangle$ and
$|qG\rangle$ interact with the target with the same total cross
section, this is why a quark cannot radiate at zero momentum
transfer and, hence, $G$ is not produced diffractively in the
forward direction. This is the general and model independent
statement. The details of this general consideration can be found in
Ref. \cite{KST-par} (Appendices A 1 and A 4). The same result is
obtained calculating Feynman graphs in Appendix B 4 of the same
paper. Unimportance of radiation between two interactions was also
demonstrated by Stan Brodsky and Paul Hoyer in Ref.~\cite{BH}.

Notice that in all these calculation one assumes that the coherence
time of radiation considerably exceeds the time interval between the
two interactions, what is fulfilled in our case, since we consider
radiation at forward rapidities.

The situation changes if the boson is radiated diffractively by a
dipole. Then the quark dipoles with or without a gauge boson have
different sizes and interact with the target differently. So the
amplitude of the diffractive gauge boson radiation from the $q{\bar
q}$ dipole is proportional to the difference between elastic
amplitudes of the two Fock components,  $|q{\bar q\rangle}$ and
$|q{\bar qG\rangle}$\cite{KPST06}, i.e.
\begin{eqnarray}
M_{\bar
qq}(\vec{b},\vec{r}_p,\vec{r},\alpha)=-2ip_1^0\,\sqrt{4\pi}\,\frac{\sqrt{1-\alpha}}{\alpha^2}\,
\Psi^{\mu}_{\gamma^*q}(\alpha,\vec{r})\left[2\mathrm{Im}\,
f_{el}(\vec{b},\vec{r_p})-2\mathrm{Im}\,
f_{el}(\vec{b},\vec{r}_p+\alpha\vec{r})\right]\label{amp-LO}
\end{eqnarray}
where $\vec r_p$ is the transverse separation of the $q\bar q$ dipole.
The partial elastic dipole-proton amplitude is normalized to the dipole cross section, which is parameterized by the following
simple ansatz \cite{GBWdip},
\begin{eqnarray}
\sigma_{\bar qq}(r_p,x)=\int
d^2b\,2\,\mathrm{Im}f_{el}(\vec{b},\vec{r}_p)=\sigma_0(1-e^{-r_p^2/R_0^2(x)}),
\label{fel}
\end{eqnarray}
where $\sigma_0=23.03\mb$; $R_0(x)=0.4\fm\times(x/x_0)^{0.144}$ and $x_0=0.003$.
This saturated form, although is oversimplified (compare with \cite{bartels}), is rather successful in description of experimental HERA
data with a reasonable accuracy. We rely on this parametrization in what follows, and the explicit form of the amplitude  $f_{el}(\vec{b},\vec{r})$, will be specified later.

The diffractive amplitude (\ref{amp-LO}), thus, occurs to be
sensitive to the large transverse separations between the projectile
quarks in the incoming proton. These distances are controlled by a
nonperturbative scale, which is one of the reasons for the breakdown
of diffractive QCD factorisation in the diffractive gauge bosons
production (for more details, see Refs.~\cite{KPST06,our-DDY}).

\section{Single diffractive cross section}

The differential cross section for the single diffractive di-lepton ($l{\bar l}$ pair in the
case of $\gamma^*,\,Z$ and $l\nu_l$ pair in the case of $W^\pm$)
production in the target rest frame can be written in terms of the
gauge boson production cross section at a given invariant mass of the di-lepton $M$. Integrating
the cross section over the solid angle of the lepton pair
and the boson transverse momentum $\vec{q}_{\perp}$ we get
for the diffractive Drell-Yan cross section \cite{our-DDY},
\begin{eqnarray}\label{DDY-cs}
\frac{d^6\sigma_{L,T}(pp\to
pl\bar{l}X)}{d^2q_{\perp}dx_1\,dM^2\,d^2\delta_{\perp}}=
\frac{\alpha_{em}}{3\pi M^2}\,\frac{d^5\sigma_{L,T}(pp\to
p\gamma^*X)}{d^2q_{\perp}dx_1\,d^2\delta_{\perp}}\,,
\end{eqnarray}
where $x_1$ is the fractional light-cone momentum of the di-lepton,
$\vec\delta_\perp$ is the transverse momentum of the recoil proton,
and $\vec{q}_{\perp}$ is the transverse momentum of the outgoing
photon (or di-lepton). Compared to our previous study, here we are
going to look at the $q_{\perp}$-dependence of the diffractive DY
cross section in a much wider range of di-lepton invariant masses
accessible at the LHC.

In the case of the diffractive production of $G=Z^0,\,W^{\pm}$
bosons, it is convenient to employ the simple and phenomenologically
successful model for the invariant mass distribution in the decay of
an unstable particle (for details, see e.g. Refs.~\cite{SMUP}) and
to present the differential cross section in the factorized form,
\begin{eqnarray}\label{WZ-cs}
\frac{d^4\sigma_{L,T}(pp\to p(G^*\to l{\bar
l},\,l\bar{\nu}_l)X)}{d^2q_{\perp}dx_1\,dM^2\,d^2\delta_{\perp}}=\mathrm{Br}(G\to
l{\bar l},\,l\nu_l)\,\rho_G(M)\,\frac{d^3\sigma_{L,T}(pp\to
pG^*X)}{d^2q_{\perp}dx_1\,d^2\delta_{\perp}}\,,
\end{eqnarray}
where $\mathrm{Br}(Z^0\to \sum_{l=e,\mu,\tau}l{\bar l})\simeq 0.101$
and $\mathrm{Br}(W^{\pm}\to \sum_{l=e,\mu,\tau}l\nu_l)\simeq 0.326$
\cite{PDG} are the leptonic branching ratios of $Z^0$ and $W^{\pm}$
bosons, and $\rho_G(M)$ is the invariant mass  distribution of the dileptons from the decay of
the gauge
boson $G$,
\begin{eqnarray}\label{rho}
\rho_G(M)=\frac{1}{\pi}\frac{M\,\Gamma_G(M)}{(M^2-m_G^2)^2+[M\,\Gamma_G(M)]^2}\,,\quad
\Gamma_G(M)/M\ll 1\,.
\end{eqnarray}
Here, $m_G$ is the fixed on-shell boson mass and $\Gamma_G(M)$ is
its total decay width defined in the standard way by substitution
$m_G\to M$, i.e.
\begin{eqnarray}\label{GammaWZ}
\Gamma_W(M)\simeq \frac{3\alpha_{em}\,M}{4\sin^2\theta_W}\,,\qquad
\Gamma_Z(M)\simeq \frac{\alpha_{em} M}{6\sin^22\theta_W}
\Bigg[\frac{160}{3}\sin^4\theta_W-40\sin^2\theta_W+21\Bigg]
\end{eqnarray}

Let us assume that the gauge boson is emitted by the quark $q_1$. As
a result of the hard emission the quark position in the impact
parameters, being initially ${\vec r}_1$, gets shifted to ${\vec
r}_1+\alpha{\vec r}$. Applying the completeness relation to the wave
function of the proton remnant in the final state
\begin{eqnarray}\nonumber
&&\sum_f\Psi_f(\vec{r}_1+\alpha
\vec{r},\vec{r}_2,\vec{r}_3;\{x_q^{1,2,...}\},\{x_g^{1,2,...}\})
\Psi^*_f(\vec{r}\,'_1+\alpha
\vec{r}\,',\vec{r}\,'_2,\vec{r}\,'_3;\{{x'}_q^{1,2,...}\},\{{x'}_g^{1,2,...}\})\\
&&\phantom{.......}=\,
\delta\bigl(\vec{r}_1-\vec{r}\,'_1+\alpha(\vec{r}-\vec{r}\,')\bigr)\delta(\vec{r}_2-\vec{r}\,'_2)
\delta(\vec{r}_3-\vec{r}\,'_3)\prod_{j}\delta(x_{q/g}^j-{x'}_{q/g}^j),
\end{eqnarray}
where $\vec r_i$, $x_{q/g}^i$ are the transverse coordinates and
fractional light-cone momenta of the valence/sea quarks and gluons,
we get the diffractive $G^*$ production cross section in the
following form \cite{KPST06,our-DDY}
\begin{eqnarray}
&&\frac{d^5\sigma_{\lambda_{G}}(pp\to
pG^*X)}{d^2q_{\perp}dx_1\,d^2\delta_{\perp}}=\frac{1}{(2\pi)^2}\frac{1}{64\pi^2}\frac{1}{x_1}\,\sum_{q=val,\,sea}\int
d^2r_1d^2r_2d^2r_3\,d^2rd^2r'\,d^2bd^2b'\,dx_{q}\prod_{i} dx_q^i dx_g^i\nonumber\\
&&\qquad\qquad\times\,\Psi^{\lambda_{G}}_{V-A}(\vec{r},\alpha,M)
\Psi^{\lambda_{G}*}_{V-A}(\vec{r}\,',\alpha,M)\,|\Psi_{i}(\vec{r}_1,
\vec{r}_2,\vec{r}_3;x_{q},\{x_q^{2,3,...}\},\{x_g^{2,3,...}\})|^2\nonumber\\
&&\qquad\qquad\times\,\Delta(\vec{r}_1,
\vec{r}_2,\vec{r}_3;\vec{b};\vec{r},\alpha)\Delta(\vec{r}_1,
\vec{r}_2,\vec{r}_3;\vec{b}\,';\vec{r}\,',\alpha)\,
e^{i\vec{\delta}_{\perp}\cdot(\vec{b}-\vec{b}\,')}\,
e^{i\vec{l}_{\perp}\cdot\alpha(\vec{r}-\vec{r}\,')}\label{eik-tot}
\end{eqnarray}
where $\Psi_{i}$ is the proton wave function, the summation is
performed over all valence/sea quarks and gluons in the proton, and
the light-cone fraction of the quark emitting the gauge boson
$x_q^1\equiv x_q$ is fixed by the external phase space variables
$x_1$ and $\alpha$ due to the momentum conservation, namely,
\begin{eqnarray}\label{xq}
x_q=\frac{x_1}{\alpha}\,,\qquad x_1=\frac{q^+}{P_1^+}
\end{eqnarray}
where $P_1$ is the 4-momentum of the projectile proton, $q$ is the
4-momentum of the produced gauge boson, and
\begin{eqnarray}\nonumber
\Delta&=&-2\mathrm{Im}\,
f_{el}(\vec{b},\vec{r}_1-\vec{r}_2)+2\mathrm{Im}\,
f_{el}(\vec{b},\vec{r}_1-\vec{r}_2+\alpha\vec{r})\\&&-2\mathrm{Im}\,
f_{el}(\vec{b},\vec{r}_1-\vec{r}_3)+2\mathrm{Im}\,
f_{el}(\vec{b},\vec{r}_1-\vec{r}_3+\alpha\vec{r})\,,
\label{eik-el}
\end{eqnarray}
is the properly normalized diffractive amplitude, where $f_{el}(\vec
b,\vec r_1-\vec r_2)$ is the partial elastic amplitude for dipole of
transverse size $r$ colliding with a proton at impact parameter $b$
to be specified below. As expected, the diffractive amplitude
$\Delta$ is proportional to the difference between elastic
amplitudes for the dipoles of slightly different sizes. This
difference is suppressed by absorptive corrections, the effect
sometimes called survival probability of large rapidity gaps.

The
amplitude Eq.~(\ref{eik-el}) is the full expression, which includes
by default the effect of absorption and does not need any extra
survival probability factor\footnote{Such a statement has already
been made in a similar analysis of the diffractive heavy flavor
production performed in Ref.~\cite{KPST07} and in our previous work
on diffractive DY study \cite{our-DDY}.}.
This can be illustrated on a simple example of elastic dipole
scattering off a potential. The dipole elastic amplitude has the
eikonal form,
\begin{equation}
\mathrm{Im}\,f_{el}(\vec{b},\vec{r}_1-\vec{r}_2)=
1-\exp\bigl[i\chi(\vec r_1)-i\chi(\vec r_2)\bigr],
\label{model}
\end{equation}
where
\begin{equation}
\chi(b)=-\int\limits_{-\infty}^\infty dz\,V(\vec b,z),
\label{eikonal}
\end{equation}
and $V(\vec b,z)$ is the potential, which depends on the impact
parameter and longitudinal coordinate, and is nearly imaginary at
high energies. The difference between elastic amplitudes with a
shifted quark position, which enters the diffractive amplitude,
reads,
\begin{equation}
\mathrm{Im}\,f_{el}(\vec{b},\vec{r}_1-\vec{r}_2+\alpha\vec r)-
\mathrm{Im}\,f_{el}(\vec{b},\vec{r}_1-\vec{r}_2)=
\exp\bigl[i\chi(\vec r_1)-i\chi(\vec r_2)\bigr]
\,\exp\bigl[i\alpha\,\vec r\cdot\vec\nabla\chi(\vec r_1)\bigr].
\label{diff}
\end{equation}
The first factor $\exp\bigl[i\chi(\vec r_1)-i\chi(\vec r_2)\bigr]$
is exactly the survival probability amplitude, which vanishes in the
black disc limit, as it should be. This proves that the cross
section Eq.~(\ref{eik-tot}) includes the effect of absorption.
Notice that usually the survival probability factor is introduced into the
diffractive cross section  probabilistically, while in
Eq.~(\ref{eik-tot}) it is treated quantum-mechanically, at the
amplitude level.

All the elastic amplitudes in Eq.~(\ref{eik-el})
implicitly depend on energy. They cannot be calculated reliably, but
but are known from phenomenology. Since large dipole sizes
$|\vec{r}_i-\vec{r}_j|\sim b\sim R_p$, $i\not=j$ ($R_p$ is the
mean proton size) are important in Eq.~(\ref{eik-el}), the Bjorken
variable $x$ is ill defined, and the collisions energy is a more
appropriate variable. A parametrization of the dipole cross section
as function of $s$ was proposed and fitted to data in
Ref.~\cite{KST-par}, and the corresponding partial dipole amplitude
is given by \cite{amir,kpss,KST-GBW-eqs}
\begin{eqnarray}\nonumber
 &&\mathrm{Im}f_{el}(\vec{b},\vec{r}_p,s,x_q)=\frac{\sigma_0(s)}{8\pi{\cal B}(s)}
 \Bigg\{\exp\Bigg[-\frac{[\vec{b}+\vec{r}_p(1-x_q)]^2}{2{\cal B}(s)}\Bigg]+
   \exp\Bigg[-\frac{[\vec{b}+\vec{r}_p x_q]^2}{2{\cal B}(s)}\Bigg]\nonumber\\
 &&-\,2\exp\Bigg[-\frac{r_p^2}{R_0^2(s)}-\frac{[\vec{b}+\vec{r}_p(1/2-x_q)]^2}
 {2{\cal B}(s)}\Bigg]\Bigg\}\,,\quad {\cal B}(s)=R_N^2(s)+R_0^2(s)/8\,,\label{KST}
\end{eqnarray}
where $x_q$ is the quark longitudinal quark fraction in the dipole
defined in Eq.~(\ref{xq}), and
\begin{eqnarray}\nonumber
 &&R_0(s)=0.88\,\mathrm{fm}\,(s_0/s)^{0.14}\,,\quad
R_N^2(s)=B_{el}^{\pi p}(s)-\frac14R_0^2(s)-\frac13\langle r_{ch}^2
\rangle_{\pi}\,,\\
 &&\qquad\qquad\qquad \sigma_0(s)=\sigma_{tot}^{\pi p}(s)
 \Big(1+\frac{3R_0^2(s)}{8\langle r_{ch}^2 \rangle_{\pi}}\Big)\,.
 \label{KST-params}
\end{eqnarray}
Here, the pion-proton total cross section is parameterized as
\cite{barnett} $\sigma_{tot}^{\pi p}(s)=23.6(s/s_0)^{0.08}$ mb,
$s_0=1000\,\GeV^2$, the mean pion radius squared is \cite{amendolia}
$\langle r_{ch}^2 \rangle_{\pi}=0.44$ fm$^2$, and the Regge
parametrization of the elastic slope $B_{el}^{\pi
p}(s)=B_0+2\alpha'_{\Pom}\ln(s/\mu^2)$, with $B_0=6\,\GeV^{-2}$,
$\alpha'_{\Pom}=0.25\,\GeV^{-2}$, and $\mu^2=1\,\GeV^2$ can be used.
We employ the $s$-dependent parametrization (\ref{KST}) in what
follows, because diffraction is essentially controlled by soft
interactions.

Finally, we parameterize the proton wave function assuming the
symmetric Gaussian shape for the spacial valence quark distributions
in the proton, as
\begin{eqnarray}\nonumber
|\Psi_i(\vec{r}_1,
\vec{r}_2,\vec{r}_3;x_{q},\{x_q^{2,3,...}\},\{x_g^{2,3,...})|^2&=&\frac{3a^2}{\pi^2}
e^{-a(r_1^2+r_2^2+r_3^2)}\rho(x_{q},\{x_q^{2,3,...}\},\{x_g^{2,3,...}\})\\
&\times&\delta(\vec{r}_1+\vec{r}_2+\vec{r}_3)\delta(1-x_q-\sum_j
x_{q/g}^j), \label{psi}
\end{eqnarray}
where sum is taken over all valence/sea quarks and gluons not
participating in the hard interaction, $x_q$ is defined in
Eq.~(\ref{xq}), $a=\langle r_{ch}^2 \rangle^{-1}$ is the inverse
proton mean charge radius squared; $\rho$ is the valence quark
distribution function in the proton. Notice that this distribution
has a low scale, so the valence quark carry the whole momentum of
the proton, while gluons and the sea are included in the constituent
valence quarks. The Gottfried sum rule based on this assumption is
know to be broken \cite{jen-chieh}, but we neglect the related
$\sim20\%$ correction.

Integrating over the fractional momenta of all partons not
participating in the hard interaction we arrive at the single
valence quark distribution in the proton, probed by the hard process
-- radiation of a heavy gauge boson,
\begin{eqnarray}\label{single-quark}
\int \prod_idx_q^idx_g^i\,\delta(1-x_q-\sum_j
x_{q/g}^j)\rho(x_{q},\{x_q^{2,3,...}\},\{x_g^{2,3,...}\})=\rho_q(x_q)\,,
\end{eqnarray}
where $q$ denotes the quark flavor emitting the gauge boson $G$ with
the fraction $x_q$ given by Eq.~(\ref{xq}). In the case of
diffractive Drell-Yan reaction \cite{our-DDY}, generalization of
the three-body proton wave function (\ref{psi}) including different
quark and antiquark flavors leads to the proton structure function as,
\begin{eqnarray}\label{SF}
\sum_q Z_q^2[\rho_q(x_q)+\rho_{\bar
q}(x_q)]=\frac{1}{x_q}F_2(x_q)\,.
\end{eqnarray}
However, in the case of diffractive $W$ and $Z$ production the
coupling factor ${\cal C}_q^Gg_{v/a,q}^G$ varies for different
(valence/sea) quark species in the proton, so one has to deal with
the original quark densities. Similar to the diffractive
DY case, in actual numerical calculations below, when summing
up the contributions of different quark flavors, we will generalize the
above approach including the sea quark and antiquark
densities in the proton at the hard scale imposed by the mass of the gauge boson. Also, the interference terms between
amplitudes corresponding to gauge boson radiated by
different valence quarks separated by large transverse distances in
the proton are strongly suppressed in the hard limit $r\ll R_0(s)$,
and are neglected.

\section{Single diffractive cross section in the forward limit}

\subsection{The two-scale approximation}

The typical hard length scale related to hard vector
boson production, $\alpha r\sim \alpha/(1-\alpha)M$, is usually
much smaller than any  hadronic scale (see, however, the next section).
Relying on smallness of the hard scale,
$\alpha r\ll R_{ij}=|\vec{r}_i-\vec{r}_j|\sim R_p$,  one can derive an
approximate analytical formulae for the diffractive cross section
(\ref{eik-tot}),
\begin{eqnarray}\label{lim}
\mathrm{Im}\,f_{el}(\vec{b},\vec{R}_{ij}+\alpha\vec r)-
\mathrm{Im}\,f_{el}(\vec{b},\vec{R}_{ij})\simeq \frac{\partial
\mathrm{Im}\,f_{el}(\vec{b},\vec{R}_{ij})}{\partial
\vec{R}_{ij}}\,\alpha\vec{r}\,,
\end{eqnarray}
For the sake of
convenience, we modify the integrals in Eq.~(\ref{eik-tot})
by introducing new variables $\vec{r}_2\to \vec{R}_{12}$
and $\vec{r}_3\to \vec{R}_{13}$, so that,
\begin{eqnarray}\label{r1int}
 \int d^2r_1d^2r_2d^2r_3\,e^{-a(r_1^2+r_2^2+r_3^2)}
 \delta(\vec{r}_1+\vec{r}_2+\vec{r}_3)=\frac19\int d^2R_{12}d^2R_{13}
 e^{-\frac{2a}{3}(R_{12}^2+R_{13}^2+\vec{R}_{12}\vec{R}_{13})}.
\end{eqnarray}
Since in the forward limit $\delta_{\perp}\to 0$ the $b$-dependence
comes only into the partial dipole amplitude $f_{el}$ defined in
Eq.~(\ref{KST}), it can be easily integrated \cite{KPST06},
\begin{eqnarray}\label{lim-int}
\int d^2b\,\frac{\partial
\mathrm{Im}\,f_{el}(\vec{b},\vec{R}_{ij})}{\partial
\vec{R}_{ij}}=\frac{\sigma_0(s)}{R_0^2(s)}\,\vec{R}_{ij}\,e^{-R_{ij}^2/R_0^2(s)}\,,
\end{eqnarray}
with the energy dependent parameters defined after Eq.~(\ref{KST}).

We see that the amplitude of diffractive gauge boson
emission in the dipole-target scattering (\ref{amp-LO}) integrated
over $\vec{b}$,
\begin{eqnarray}\label{fac-break}
\int d^2b\, M_{qq}(\vec{b},\vec{R}_{ij},\vec{r},\alpha)\propto
\alpha\frac{\sigma_0(s)}{R_0^2(s)}\,(\vec{r}\cdot\vec{R}_{ij})\,e^{-R_{ij}^2/R_0^2(s)},
\end{eqnarray}
is proportional to the product of the hard scale $r\sim
1/(1-\alpha)M$ and the soft hadronic scale $R_{ij}\sim R_0\sim
1/\Lambda_{QCD}$. This means that the single diffractive cross
section depends on the hard scale as $\sigma_{sd}\sim r^2\sim
1/M^2$.

It is well-known that the cross section of diffractive
Deep-Inelastic-Scattering (DDIS) $\sigma_{\rm{DDIS}}\sim r^4$ is
dominated essentially by soft fluctuations at large $r$ (for more
details, see e.g. Ref.~\cite{povh}), as correctly predicted by the
diffractive (Ingelman-Schlein) QCD factorisation. This happens since
the end-point $q{\bar q}$ dipole fluctuations, driving the cross
section at $\alpha\to0$ or 1, have no hard scale dependence for
light quarks $m_q\ll Q^2$. In this case, the $Q^2$-dependence comes
only into their weight as $\sim 1/Q^2$, even though it is of the
higher twist nature.

In opposite, the single diffractive gauge bosons production cross
section behaves as $\sim \vec{r}\cdot \vec{R}$, soft and hard
fluctuations contribute in this process on the same footing, and
their interplay does not depend on the hard scale, similar to the
inclusive gauge bosons production. Hence, the forward diffractive
abelian radiation turns out to be of the leading twist nature, and
the diffractive-to-inclusive production cross sections ratio can
depend on the hard scale only weakly through the $x$-dependence of
the saturation scale, or more precisely $R_0(x_2)$, and can only
increase (see below).

However, if one uses the conventional diffractive factorisation
scheme \cite{ISh} the single diffractive cross section, similarly to
the DDIS process, one does not find any soft-hard interplay as
observed above, and the cross section turns out to behave as $\sim
r^4$, providing the higher twist nature of the single diffractive
process. Correspondingly, this strongly affects the $M^2$-dependence
of the diffractive-to-inclusive boson production cross sections
ratio, such that it decreases with $M^2$, as opposite to our
observation above.

Therefore, the fundamental interplay between the hard and soft
interactions in the forward diffractive Abelian radiation is {\it
the major reason for the diffractive QCD factorisation breaking}
leading to quite unusual features of the corresponding observables
(for a similar discussion in the diffractive DY, see
Refs.~\cite{KPST06,our-DDY}). As we have emphasized above, this
interplay is absent in the DDIS and in diffractive QCD
factorisation-based approaches to the diffractive DY (see e.g.
Ref.~\cite{Szczurek}) leading to the energy and scale dependence of
the corresponding cross section which is completely opposite to the
one predicted above by the Color Dipole model.

Further, the integrations over $\vec{R}_{12}$ and $\vec{R}_{13}$ can
be performed analytically leading to the diffractive cross section
(\ref{eik-tot}) in the forward limit $\delta_{\perp}\to 0$,
\begin{eqnarray} \label{eik-tot-s1}
&&\frac{d^4\sigma_{\lambda_{G}}(pp\to
p\,G^*X)}{d^2q_{\perp}dx_1\,d\delta_{\perp}^2}\Big|_{\delta_{\perp}=0}=
\frac{a^2}{24\pi^3}\frac{\sigma_0^2(s)}{R_0^4(s)}\,
\frac{1}{A_2}\Big[\frac{2}{(A_2-4A_1)^2}+\frac{A_2^2}{(A_2^2-4A_3^2)^2}\Big]\times\\
&&\sum_q \int_{x_1}^1 d\alpha\,
\Big[\rho_q\Big(\frac{x_1}{\alpha}\Big)+
\rho_{\bar{q}}\Big(\frac{x_1}{\alpha}\Big)\Big]\int
d^2rd^2r'\,(\vec{r}\cdot\vec{r}\,')\,\Psi^{\lambda_{G}}_{V-A}(\vec{r},\alpha,M)
\Psi^{\lambda_{G}*}_{V-A}(\vec{r}\,',\alpha,M)\,
e^{i\vec{q}_{\perp}\cdot(\vec{r}-\vec{r}\,')},\nonumber
\end{eqnarray}
where
\begin{eqnarray}\label{Aj}
A_1=\frac{2a}{3}+\frac{2}{R_0^2(s)}\,,\qquad
A_2=\frac{2a}{3}\,,\qquad A_3=\frac{2a}{3}+\frac{1}{R_0^2(s)}\,.
\end{eqnarray}

Assuming gaussian $\delta_{\perp}$ dependence of the cross section,
the $\delta_{\perp}$ integrated and forward cross sections are related as,
\begin{eqnarray}
\frac{d\sigma(pp\to
p\,G^*X)}{d^2q_{\perp}dx_1}=\frac{1}{B_{sd}(s)}\frac{d^3\sigma(pp\to
p\,G^*X)}{d^2q_{\perp}dx_1\,d\delta_{\perp}^2}\Big|_{\delta_{\perp}=0}.
\end{eqnarray}
The slope of the single-diffractive cross section, $B_{sd}(s)\simeq
\langle r_{ch}^2 \rangle/3+2\alpha'_{\Pom}\ln(s/s_0)$, is similar to
the one measured in diffractive DIS. In the next section we will
explicitly derive the diffractive slope from the explicit
parameterisation for the partial dipole amplitude (\ref{KST}).

Finally, one can explicitly calculate the remaining integrations in
the transverse plane over $\vec{r}$ and $\vec{r}\,'$ by means of the
following Fourier transforms
\begin{eqnarray}\nonumber
&&J_1(q_{\perp},\eta)\equiv\int
d^2rd^2r'\,(\vec{r}\cdot\vec{r}\,')\,K_0(\eta r)K_0(\eta r')\,
e^{i\vec{q}_{\perp}\cdot(\vec{r}-\vec{r}\,')}=16\pi^2\frac{q_{\perp}^2}{(\eta^2+q_{\perp}^2)^4}\,,\\
&&J_2(q_{\perp},\eta)\equiv\int
d^2rd^2r'\,\frac{(\vec{r}\cdot\vec{r}\,')^2}{rr'}\,K_1(\eta
r)K_1(\eta r')\,
e^{i\vec{q}_{\perp}\cdot(\vec{r}-\vec{r}\,')}=8\pi^2\frac{\eta^4+q_{\perp}^4}{\eta^2(\eta^2+q_{\perp}^2)^4}\,.
\end{eqnarray}
And we arrive at  the following expressions for the cross section of
transversely and longitudinally polarized gauge boson production,
respectively,
\begin{eqnarray} \label{eik-tot-T}
&&\frac{d^4\sigma_{T}(pp\to
p\,G^*X)}{d^2q_{\perp}dx_1}=\frac{1}{B_{sd}(s)}
\frac{a^2}{24\pi^3}\frac{\sigma_0^2(s)}{R_0^4(s)}\,
\frac{1}{A_2}\Big[\frac{2}{(A_2-4A_1)^2}+\frac{A_2^2}{(A_2^2-4A_3^2)^2}\Big]\times\\
&&\sum_q \frac{({\cal C}^G_q)^2}{2\pi^2}\,\int_{x_1}^1 d\alpha\,
\Big[\rho_q\Big(\frac{x_1}{\alpha}\Big)+
\rho_{\bar{q}}\Big(\frac{x_1}{\alpha}\Big)\Big]\Big\{m_q^2\alpha^2
\Big[(g_{v,q}^G)^2 \alpha^2 + (g_{a,q}^G)^2 (2-\alpha)^2\Big] J_1 +
\nonumber
\\ && \Big[(g_{v,q}^G)^2 + (g_{a,q}^G)^2\Big]
\left[1+\left(1-\alpha\right)^2\right]\eta^2 J_2\Big\}\,;\nonumber
\end{eqnarray}
\begin{eqnarray} \label{eik-tot-L}
&&\frac{d^4\sigma_{L}(pp\to
p\,G^*X)}{d^2q_{\perp}dx_1}=\frac{1}{B_{sd}(s)}
\frac{a^2}{24\pi^3}\frac{\sigma_0^2(s)}{R_0^4(s)}\,
\frac{1}{A_2}\Big[\frac{2}{(A_2-4A_1)^2}+\frac{A_2^2}{(A_2^2-4A_3^2)^2}\Big]\times\\
&&\sum_q \frac{({\cal C}^G_q)^2}{\pi^2}\,\int_{x_1}^1 d\alpha\,
\Big[\rho_q\Big(\frac{x_1}{\alpha}\Big)+
\rho_{\bar{q}}\Big(\frac{x_1}{\alpha}\Big)\Big]\Big\{\Big[(g_{v,q}^G)^2
M^2 \left(1-\alpha\right)^2 + (g_{a,q}^G)^2\frac{\eta^4}{M^2} \Big]
J_1 + \nonumber \\&& (g_{a,q}^G)^2\alpha^2m_q^2
\frac{\eta^2}{M^2}\,J_2\Big\}\,.\nonumber
\end{eqnarray}
These expressions for the differential distributions in the transverse
momentum of the produced gauge bosons  allow us to perform
$\vec{q}_{\perp}$-integration via the substitution,
\begin{eqnarray*}
&&J_1(q_{\perp},\eta)\to I_1(\eta)\equiv\int
d^2q_{\perp}\,J_1(q_{\perp},\eta)=\frac{8\pi^3}{3\eta^4}\,,\\
&&J_2(q_{\perp},\eta)\to I_2(\eta)\equiv\int
d^2q_{\perp}\,J_2(q_{\perp},\eta)=\frac{16\pi^3}{3\eta^4}\,.
\end{eqnarray*}
The rest of integrations over $\alpha$ and $x_1$ can be done numerically.

\subsection{Asymptotic behavior of diffraction}

In most of theoretical models the partial elastic amplitude of
hadron scattering is expected to reach the Froissart regime, which
corresponds to saturation of unitarity for the amplitudes of all
Fock states within a disc in impact parameter plane, with radius
rising as $\ln s$. In this case diffraction vanishes everywhere,
except the periphery of the disc, so the the fraction of diffractive
cross section is expected to fall with energy as
$\sigma_{diff}/\sigma_{tot}\propto1/\ln s$ \cite{KPSdiff}.

However, Eqs.~(\ref{eik-tot-s1}), (\ref{Aj}) at $s\to\infty$, when
$R_0(s)\to0$, lead to the forward diffractive cross section
proportional to $\sigma_{tot}^2(s)$. So the $b$-integrated
diffractive cross section behaves like the elastic and total cross
sections, contrary to the above expectations.

To trace the origin of this problem we should recheck our starting
assumptions. So far we assumed that the hard length scale
$r\sim1/M$, which is indeed tiny, is much smaller than any other
soft hadronic scale. However, the hierarchy of scales is expected to
change at asymptotically high energies. This is related to the
rising energy dependence of the saturation scale in the proton,
$Q_s=1/R_0(s)=0.086\GeV\times (s/1\GeV^2)^{0.14}$, where we rely on
the parametrization Eq.~(\ref{KST-params}). Although this power
energy dependence is rather steep, the saturation scale reaches the
gauge boson mass at the energy $s\approx 3\times 10^9\GeV^2$, which
is much higher than that of the LHC. Therefore, in the energy range
of LHC the hard length scale $\sim 1/M$ remains the shortest in the
process, and the expansion Eq.~(\ref{lim}) can be employed. Indeed,
we perform an exact numerical calculation of the cross section
Eq.~(\ref{eik-tot}) and found no sizable deviation from
Eq.~(\ref{eik-tot-s1}).

Apparently the root of the problem the approximation of $1/M\ll
R_0(s)$, which breaks at very large $s$. To demonstrate analytically
that the fractional diffractive cross section (\ref{eik-tot})
vanishes at asymptotic energies. Let us consider a simple example of
radiation
 os a heavy gauge boson by a dipole of size $R$. In this case the forward
 diffractive amplitude, i.e. the amplitude Eq.~(\ref{eik-el})
 integrated over $b$ takes the form ($\alpha=1$),
 \beq
 A(\vec R,\vec r,s,\delta_\perp=0)=\sigma_{\bar qq}(\vec R+ \vec r,s)-
 \sigma_{\bar qq}(\vec R,s)=
 \sigma_0(s)\left[e^{-R^2/R_0^2(s)}-e^{-(\vec R+ \vec r)^2/R_0^2(s)}\right].
 \label{Delta}
 \eeq
 This amplitude should be averaged over the $R$- and $r$-distributions,
 \beqn
 A(s,\delta_\perp=0)&=&\int d^2R\,d^2r\,|\Psi_h(R)|^2|\Psi_{qG}(r)|^2\Delta(\vec R,\vec r,s) \nonumber\\ &=&
 \pi R_0^2(s)\sigma_0(s)\left[|\Psi_h(0)|^2-\int d^2r\,|\Psi_h(r)|^2|\Psi_{qG}(r)|^2\right]
 \nonumber\\ &=& R_0^2(s)\,
 \frac{\sigma_0(s)\,\la r^2\ra/\la R^2\ra}{\la R^2\ra+\la r^2\ra}\approx
 R_0^2(s)\sigma_0(s)\,\frac{\la r^2\ra}{\la R^2\ra^2}
 \label{average}
 \eeqn
Here we we assumed the energy to be sufficiently high, so
$R_0(s)\to0$ and $R_0^2(s)\ll \la r^2\ra$. We also assumed that the
distribution function of the dipole $\Psi_h(R)$ and of the $q-G$
fluctuation, $\Psi_{qG}(r)$ have gaussian form with averaged values
$\la R^2\ra$ and $\la r^2\ra$ respectively.

According to the parametrization (\ref{KST-params}) the ratio of
forward diffractive amplitude Eq.~(\ref{average}) to the forward
elastic amplitude vanishes as, \beq \frac{ A(s,\delta_\perp=0)}{
A_{el}(s,\delta_\perp=0)}\approx R_0^2(s)\,\frac{2\la r^2\ra}{\la
R^2\ra^2}\propto s^{-0.28}. \label{diff/el} \eeq I fact, the
parametrization (\ref{KST-params}) fitted to available data, is not
expected to be valid at asymptotically high energies. The saturation
scale corresponding to transition from the linear BFKL evolution at
$Q>Q_s$ to the nonlinear saturation regime at $Q<Q_s$ rises with
energy as $Q_s=1/R_0(s)\propto \exp[const\times\sqrt{\ln s}]$
\cite{al}. So it is steeply falling with energy, also slower than in
(\ref{diff/el}).

Thus, the diffractive amplitude vanishes at very high, currently
unreachable energies, while within the available energy range the
expression (\ref{eik-tot-s1}) is sufficiently accurate.

\section{Diffractive vs inclusive production of gauge bosons}

The dipole description of inclusive Gauge boson production can be
obtained generalizing what is known for  the inclusive Drell-Yan
process \cite{deriv1,deriv2,KRT00}. The cross section of inclusive
production of a virtual gauge boson $G^*$ with mass $M$ and
transverse momentum $q_{\perp}$ has the form,
\begin{eqnarray}\label{inclGB}
&&\frac{d^4\sigma_{\lambda_G}(pp\to
G^*X)}{d^2q_{\perp}\,dx_1}=\frac{1}{(2\pi)^2}\sum_q \int_{x_1}^1
\frac{d\alpha}{\alpha^2}\, \Big[\rho_q\Big(\frac{x_1}{\alpha}\Big)+
\rho_{\bar{q}}\Big(\frac{x_1}{\alpha}\Big)\Big]\times\\&& \int
d^2rd^2r'\,\frac12\Big\{\sigma(\alpha r)+\sigma(\alpha
r')-\sigma(\alpha|\vec r-\vec
r\,'|)\Big\}\,\Psi^{\lambda_{G}}_{V-A}(\vec{r},\alpha,M)
\Psi^{\lambda_{G}*}_{V-A}(\vec{r}\,',\alpha,M)\,
e^{i\vec{q}_{\perp}\cdot(\vec{r}-\vec{r}\,')}\,.\nonumber
\end{eqnarray}

The principal difference of the inclusive gauge boson production
from the diffractive one is in the typical size of the dipoles
involved in the scattering. As is seen from e.g. Eqs.~(\ref{fel}),
(\ref{eik-el}), the diffractive scattering is dominated by large
dipoles scattering at the hadronic scale, with the transverse size
$r_p=R_{ij}\sim R_0$ (soft scattering), whereas the inclusive
production cross section (\ref{inclGB}) is totally driven by
small-size dipoles scattering with $r_p=\alpha r \ll R_0$ (hard
scattering). Therefore, different parameterizations for the dipole
cross sections must be used -- in the diffractive case above we have
adopted the KST parametrization for the dipole cross section (or the
partial amplitude (\ref{KST})) with $s$-dependent parameters
introduced in Eq.~(\ref{KST-params}) \cite{KST-GBW-eqs,KST-par},
whereas in the inclusive production case the Bjorken $x$-dependent
GBW parametrization Eq.~(\ref{fel}) \cite{GBWdip} is better
justified:
\begin{eqnarray}
\bar{\sigma}_0=23.03\,{\rm mb}\,,\quad R_0\equiv
\bar{R}_0(x_2)=0.4\,\mathrm{fm}\times(x_2/x_0)^{0.144}\,, \quad
x_0=3.04\times10^{-4}\,, \label{GBW-params}
\end{eqnarray}
where $x_2=q^-/P_2^-$, with $P_2$ being the 4-momentum of the target
proton.

In the leading regime of $\alpha r,\,\alpha r' \ll R_0$.
\begin{eqnarray}
\frac12\Big\{\sigma(\alpha r)+\sigma(\alpha r')-\sigma(\alpha|\vec
r-\vec r\,'|)\Big\}\simeq
\frac{\alpha^2\bar{\sigma}_0}{\bar{R}_0^2(x_2)}\,(\vec{r}\cdot\vec{r}\,'),
\end{eqnarray}
so the inclusive gauge boson production cross section at forward
rapidities ($x_1\gg x_2$) reads,
\begin{eqnarray}\label{inclGB-1}
&&\frac{d^4\sigma_{\lambda_G}(pp\to
G^*X)}{d^2q_{\perp}\,dx_1}=\frac{1}{(2\pi)^2}\frac{\bar{\sigma}_0}{\bar{R}_0^2(x_2)}\,
\sum_q \int_{x_1}^1 d\alpha\,
\Big[\rho_q\Big(\frac{x_1}{\alpha}\Big)+
\rho_{\bar{q}}\Big(\frac{x_1}{\alpha}\Big)\Big]\times\\&& \int
d^2rd^2r'\,(\vec{r}\cdot\vec{r}\,')\,\Psi^{\lambda_{G}}_{V-A}(\vec{r},\alpha,M)
\Psi^{\lambda_{G}*}_{V-A}(\vec{r}\,',\alpha,M)\,
e^{i\vec{q}_{\perp}\cdot(\vec{r}-\vec{r}\,')}\,.\nonumber
\end{eqnarray}
We observe that the integrals over $\alpha$ and
$\vec{r},\,\vec{r}\,'$ have the same form as in the
diffractive cross section Eq.~(\ref{eik-tot-s1}).

The $M$-dependence of the differential cross sections for
di-lepton inclusive production via an intermediate photon $\gamma^*$ or a gauge
boson $G^*$ can be presented similar to the diffractive case, as
  \cite{KRT00},
\begin{eqnarray}\label{incl-cs}
&&\qquad\qquad\quad\frac{d\sigma_{\lambda_{\gamma}}(pp\to
(\gamma^*\to
l\bar{l})X)}{d^2q_{\perp}dx_1\,dM^2}=\frac{\alpha_{em}}{3\pi
M^2}\,\frac{d\sigma(pp\to \gamma^*X)}{d^2q_{\perp}dx_1}\,;\\
&&\frac{d\sigma(pp\to (G^*\to
l\bar{l},\,l\bar{\nu}_l)X)}{d^2q_{\perp}dx_1\,dM^2}=
\mathrm{Br}(G\to l{\bar l},\,l\nu_l)\,\rho_G(M)\,\frac{d\sigma(pp\to
G^*X)}{d^2q_{\perp}dx_1}\,, \nonumber
\end{eqnarray}
where the resonance mass distribution $\rho_G(M)$
is given by Eq.~(\ref{rho}).

Eventually, we arrive at a simple form for the ratio of the
diffractive and inclusive cross sections for
di-lepton production,
\begin{eqnarray}\label{ratDDincl}
\frac{d\sigma^{sd}_{\lambda_G}/d^2q_{\perp}\,dx_1\,dM^2}
{d\sigma^{incl}_{\lambda_G}/d^2q_{\perp}dx_1\,dM^2}=
\frac{a^2}{6\pi}\frac{\bar{R}_0^2
(M_{\perp}^2/x_1s)}{B_{sd}(s)\,\bar{\sigma}_0}\frac{\sigma_0^2(s)}{R_0^4(s)}\,
\frac{1}{A_2}\Big[\frac{2}{(A_2-4A_1)^2}+\frac{A_2^2}{(A_2^2-4A_3^2)^2}\Big]
\end{eqnarray}
where functions $A_{1,2,3}$ were defined in Eq.~(\ref{Aj}), and
fraction $x_2$ is explicitly given in terms of other kinematic
variables in Eq.~(\ref{Aj}).

It turns out that the ratio (\ref{ratDDincl}) does not depend either
on the type of the intermediate boson, or on its helicity
$\lambda_G$. To a good approximation, it is controlled mainly by
soft interaction dynamics, in terms of the soft parameters only
$\bar{R_0},\,R_0,\,\bar{\sigma}_0$ and $\sigma_0$. A slow dependence
of these parameters on the collision energy $s$, the hard scale
$M^2$ and the boson transverse momentum $q_{\perp}$ completely
determines such dependence of the diffractive-to-inclusive
production ratio. A measurement of the $M^2$ (or $q_{\perp}$)
dependence of this ratio would allow to probe the $x$-evolution of
the saturation scale, as well as to constrain its energy dependence.
Hence, such a quantity is a very useful probe for the underlined QCD
diffractive mechanism and the saturation phenomenon, and will be
quantified based on existing KST/GBW parameterizations in the next
section.

\section{Breakdown of diffractive factorization}

It is instructive to trace the origin of QCD factorisation in
inclusive processes within the dipole description. The $1/Q^2$
dependence of the DIS cross section at small $x$ originates from two
different sources. Most of the $\bar qq$ fluctuations of a virtual
photon have a small size, $r^2\sim 1/Q^2$, except the endpoint
(aligned jet) configurations with $\alpha\to 0,\ 1$.  The latter
have large hadronic size and cross section, but their weight is
small $\sim1/Q^2$. Thus, both contributions to the cross section
behave as $1/Q^2$.

Similarly, in the Drell-Yan process of radiation of the heavy photon
with fractional momentum $x_1$ the mean size of the artificial
dipole \cite{deriv1} has small size $r\sim 1/(1-\alpha)Q$, except
the endpoint configurations with $\alpha\to1$. Like in DIS, $\alpha$
is not an observable, but is integrated from $x_1$ to 1 (see
Eq.~(\ref{inclGB})). This similarity, which reflects the
factorisation relation between the two processes, also demonstrates
its limitations. At large $x_1\to1$ the inclusive Drell-Yan reaction
is fully dominated by the soft component and factorisation breaks
down.

For the diffractive channels the factorisation relation breaks down
at any $x_1$. DIS diffraction is fully dominated by the soft
dynamics, since the probability of endpoint configurations is still
the same, $\propto 1/Q^2$, while the cross section is enhanced by a
factor of $Q^4$ compared with the hard component \cite{povh}.
However, the mechanism of Drell-Yan diffraction is quite different
\cite{KPST06,our-DDY}, apparently breaking the factorisation
relation. It comes from the hard-soft interference, which imitates a
leading twist throughout the whole range of $x_1$. Again, like in
inclusive process, the hard and soft (endpoint) components make
comparable contributions to the diffractive Drell-Yan cross
sections. Of course, this is true for other gauge bosons as well.

Although involvement of large distances in diffractive heavy boson
production is in obvious contradiction with factorisation of hard
and soft scales, an observable manifestation of that is not trivial.
Indeed, the cross section of a hard process $q\bar q\to l\bar l$
with the sea quark density in the Pomeron measured in diffractive
DIS, although it is a higher twist, imitates the leading twist scale
dependence. Nevertheless, the predicted scale and energy
dependencies are quite different, as we demonstrate in the next
Section~\ref{numerics}. Notice that a much more pronounced breakdown
of diffractive factorisation was previously found in
Ref.~\cite{KPST07} for the case when the hard scale is imposed by
the mass of a heavy flavor.

\section{Numerical results}\label{numerics}

We now turn to a discussion of the numerical results for the most
important observables. First of all, we are interested in the di-lepton ({\it
\`a la} Drell-Yan pair) production channel as the simplest one. Although quark
production channel could also be of interest, this case will be
considered elsewhere.
\begin{figure*} [!h]
\begin{minipage}{0.49\textwidth}
 \centerline{\includegraphics[width=1.0\textwidth]{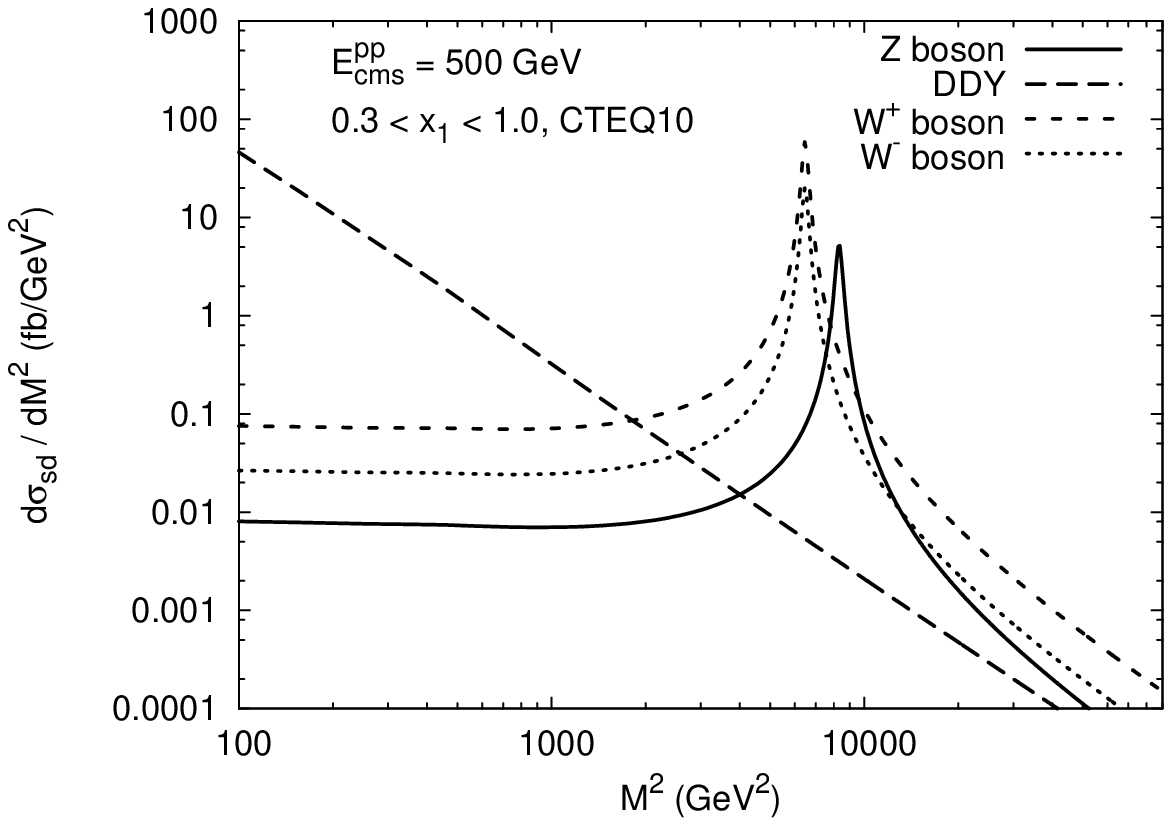}}
\end{minipage}
\hspace{0.5cm}
\begin{minipage}{0.46\textwidth}
 \centerline{\includegraphics[width=1.0\textwidth]{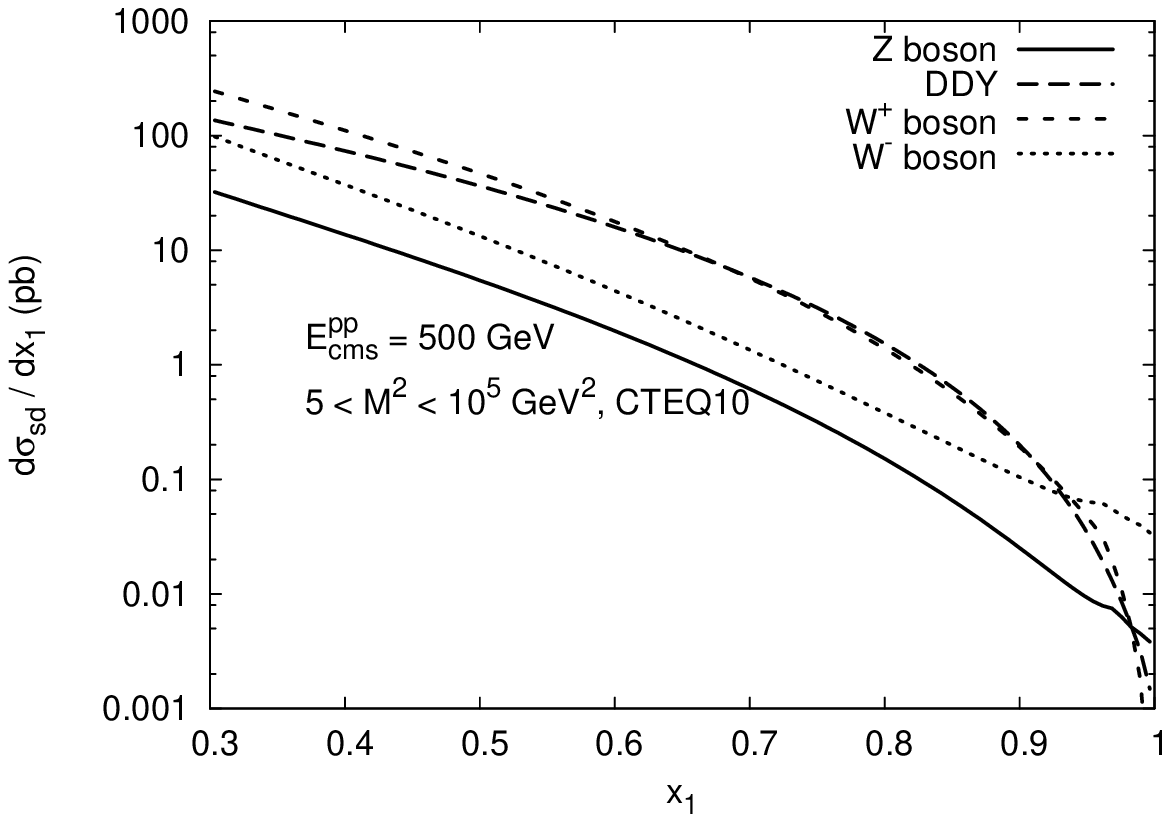}}
\end{minipage}
   \caption{
\small Diffractive gauge boson production cross section as function
of di-lepton invariant mass squared $M^2$ (left panel) and boson
fractional light-cone momentum $x_1$ (right panel) in $pp$
collisions at the RHIC  energy $\sqrt{s}=500$ GeV. Solid,
long-dashed, dashed and dotted curves correspond to $Z$, $\gamma^*$,
$W^+$ and $W^-$ bosons, respectively. CTEQ10 PDF parametrization
\cite{CTEQ10} is used here.}
 \label{fig:CS-RHIC}
\end{figure*}

\begin{figure*}[!b]
\begin{minipage}{0.49\textwidth}
 \centerline{\includegraphics[width=1.0\textwidth]{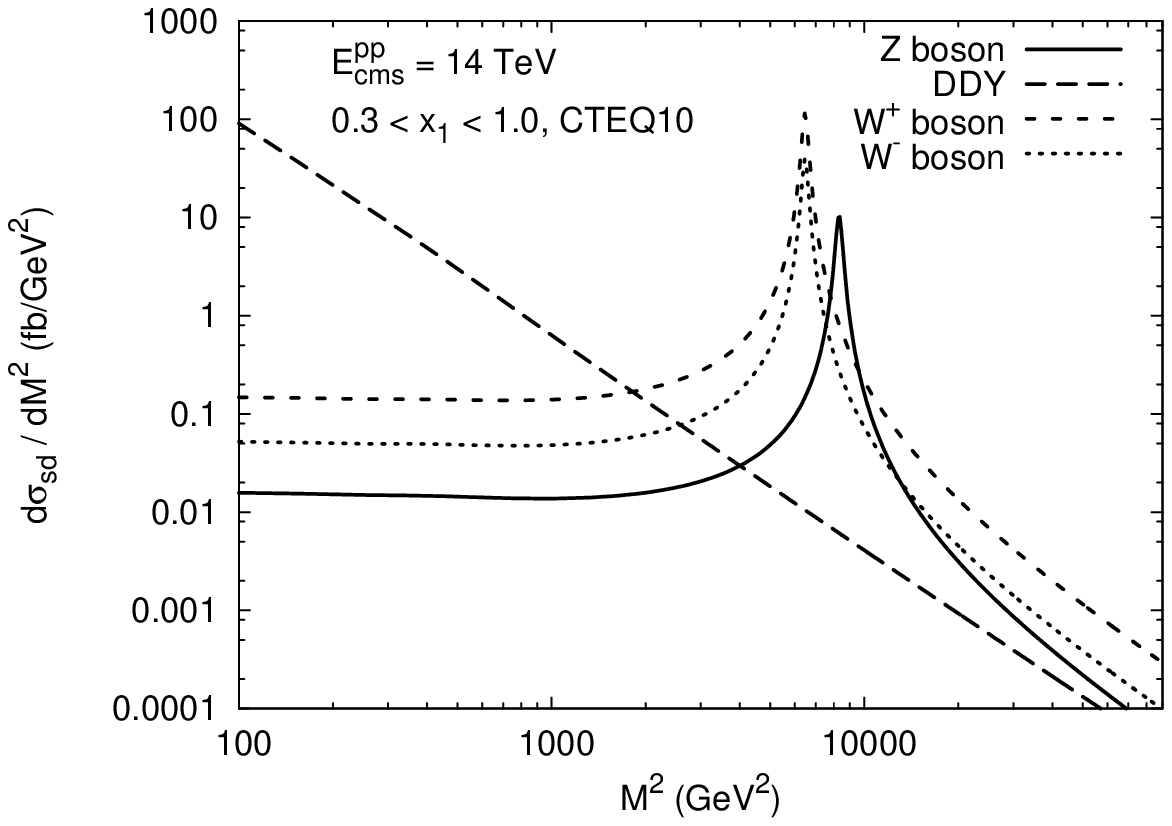}}
\end{minipage}
\hspace{0.5cm}
\begin{minipage}{0.46\textwidth}
 \centerline{\includegraphics[width=1.0\textwidth]{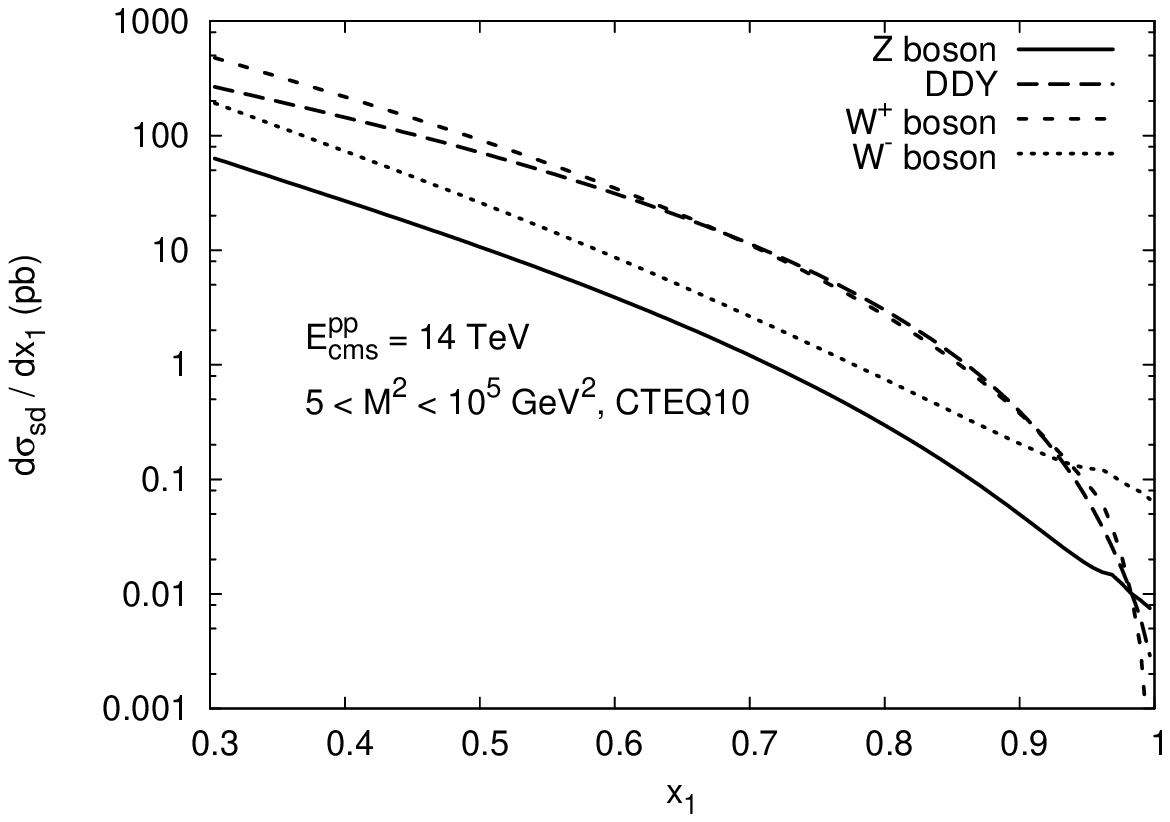}}
\end{minipage}
   \caption{
\small The same as in Fig.~\ref{fig:CS-RHIC}, but for the LHC energy
$\sqrt{s}=14$ TeV.}
 \label{fig:CS-LHC}
\end{figure*}

In Fig.~\ref{fig:CS-RHIC} (for RHIC energy $\sqrt{s}=500$ GeV) and
Fig.~\ref{fig:CS-LHC} (for LHC energy $\sqrt{s}=14$ TeV) we present
the single diffractive cross sections for $Z^0,\,\gamma^*$
(diffractive DY) and $W^{\pm}$ bosons production, differential in
the di-lepton mass squared $d\sigma_{sd}/dM^2$ (left panels) and its
longitudinal momentum fraction, $d\sigma_{sd}/dx_1$ (right panels).
These plots  do not reflect particular detector constraints -- a
thorough analysis including detector acceptances and cuts has to be
done separately. The $M^2$ distributions here are integrated over
the {\it ad hoc} interval of fractional boson momentum $0.3<x_1<1$,
corresponding to the forward rapidity region (at not extremely large
masses). Then the mass distribution is integrated over the
potentially interesting invariant mass interval $5<M^2<10^5$
GeV$^2$, and can be easily converted into (pseudo)rapidity ones
widely used in experimental studies, if necessary.

The $M^2$ distributions of the $Z^0$ and $W^{\pm}$ bosons clearly
demonstrate their resonant behavior, and in the resonant region
significantly exceed the corresponding diffractive Drell-Yan
component; only for very low masses the $\gamma^*$ contribution
becomes important (left panels). For $x_1$ distribution, when
integrated over low mass and resonant regions, diffractive $W^+$ and
$\gamma^*$ components become comparable to each other, both in
shapes and values, whereas the $W^-$ and, especially, $Z$-boson
production cross section are noticeably lower (right panels). Quite
naturally, the $W^-$ cross section is (in analogy with the
well-known inclusive $W^{\pm}$ production) smaller than the $W^+$
one due to differences in valence $u$- and $d$-quark densities
(dominating over sea quarks at large $x_q$) in the proton, the
bosons couple to. So the precise measurement of differences in
forward diffractive $W^+$ and $W^-$ rates would allow to constrain
quark content of the proton at large $x_q\equiv x_1/\alpha$. In
Fig.~\ref{fig:CS-RHIC} and \ref{fig:CS-LHC}, and in all calculation
below we have used the most recent CTEQ10 valence/sea quark PDFs
parametrization \cite{CTEQ10}, if not declared otherwise.
\begin{figure}[!h]
\begin{minipage}{0.49\textwidth}
 \centerline{\includegraphics[width=1.0\textwidth]{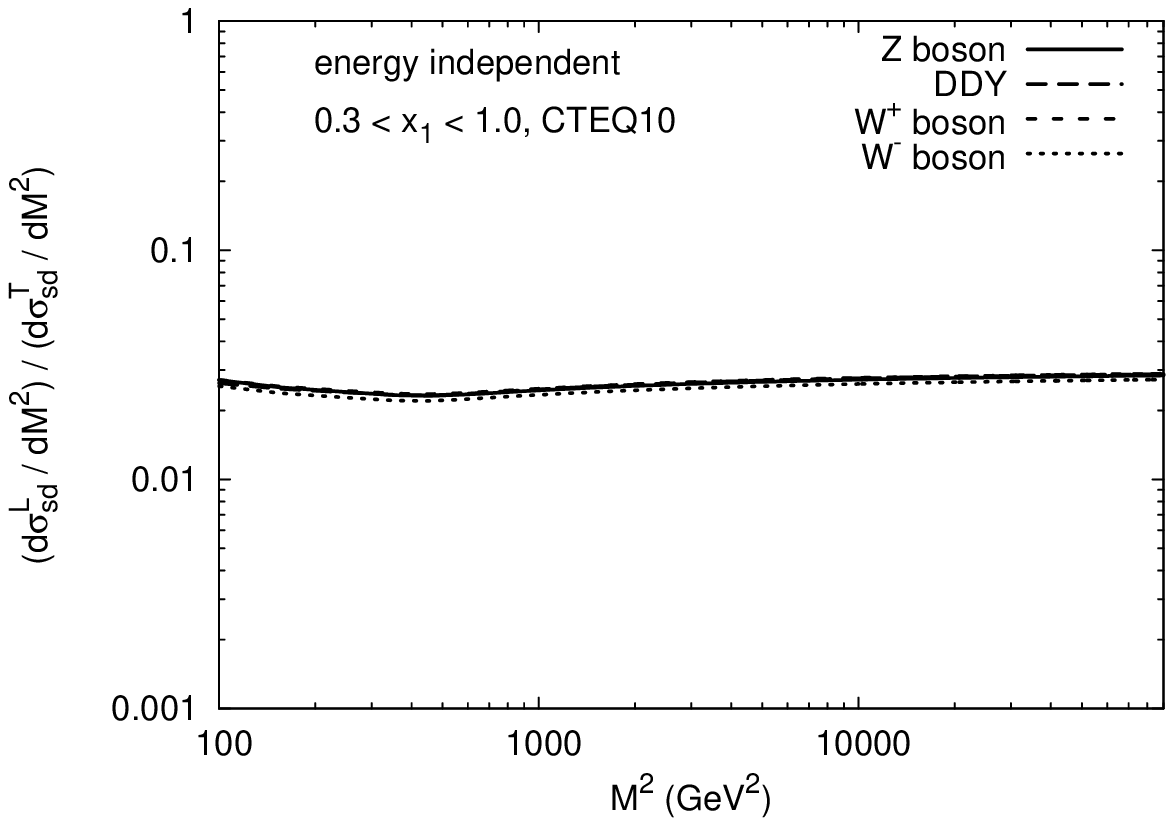}}
\end{minipage}
\hspace{0.5cm}
\begin{minipage}{0.46\textwidth}
 \centerline{\includegraphics[width=1.0\textwidth]{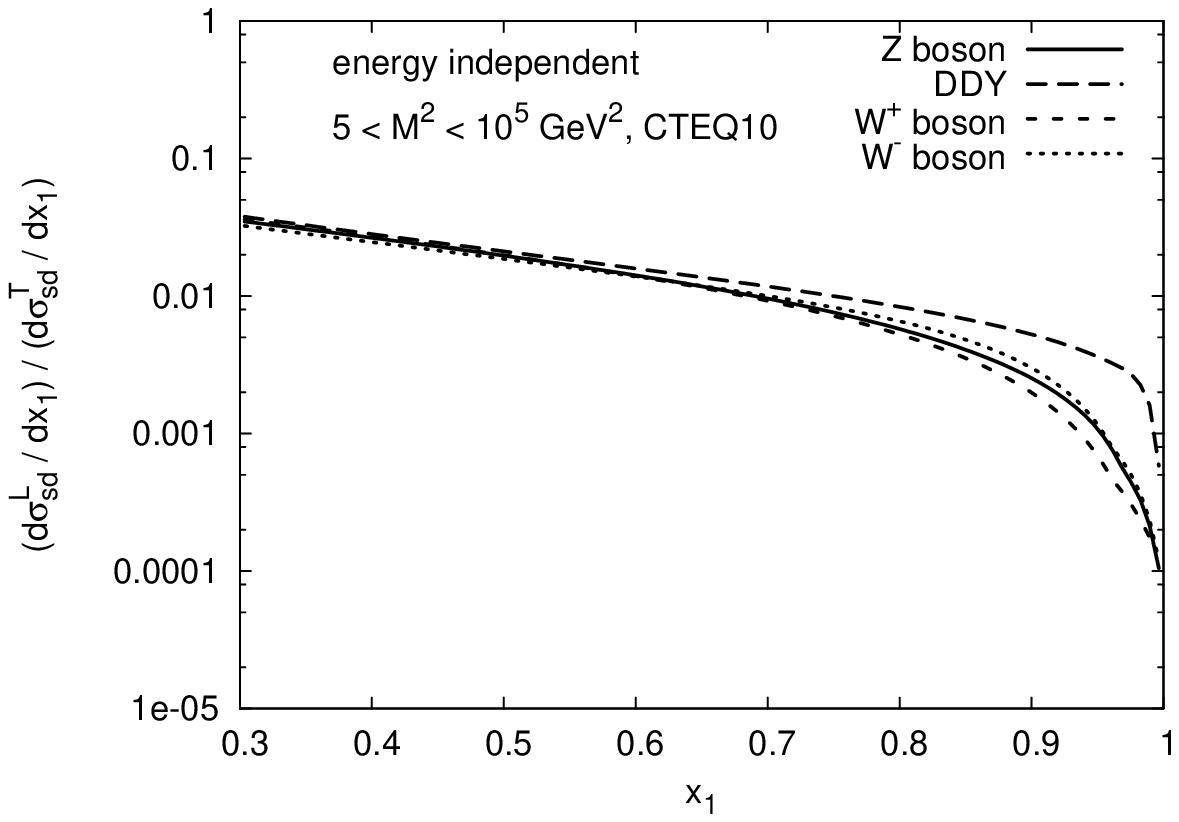}}
\end{minipage}
   \caption{
\small The ratio of the cross sections of longitudinally (L) to
transversely (T) polarized gauge bosons, as function of the
di-lepton invariant mass squared $M^2$ (left panel), and boson
fractional light-cone momentum $x_1$ (right panel). Solid,
long-dashed, dashed and dotted curves correspond to $Z$, $\gamma^*$,
$W^+$ and $W^-$ bosons, respectively. CTEQ10 PDF parametrization
\cite{CTEQ10} is used here. The ratio depends on $pp$ c.m. energy
only slightly, by a few percents ({\it cf.} Ref.~\cite{our-DDY})
over a vast multi-TeV interval, so we neglect it here.}
 \label{fig:LvsT}
\end{figure}

In Fig.~\ref{fig:LvsT} we show the ratio of the longitudinal (L) to
transverse (T) gauge boson polarization contributions to the
diffractive production cross section. This ratio is presented
differentially as function of lepton-pair invariant mass squared
$(\sigma^L_{sd}/dM^2)/(\sigma^T_{sd}/dM^2)$ (left panel) and gauge
boson fractional momentum
$(\sigma^L_{sd}/dx_1)/(\sigma^T_{sd}/dx_1)$ (right panel). We see
that the diffractive gauge bosons production process is always
dominated by radiation of transversely polarized lepton pairs. The
ratio $\sigma_L/\sigma_T$ only slightly depends on $M^2$ and even
less on $pp$ c.m. energy $\sqrt{s}$, so it can be considered as
energy independent (which, in fact, can be already seen from
approximate formulae (\ref{eik-tot-T}) and (\ref{eik-tot-L})). The
longitudinal bosons polarization roughly amounts to 10 \% at
$x_1\sim 0.5$ and then steeply falls down at large $x_1\to1$
asymptotically approaching relativistic (massless) bosons case
given, due to the gauge invariance, by the transverse polarisation
only. Such a behavior turns out to be the same as in the inclusive
Drell-Yan process \cite{KPST06}. At smaller $x_1\lesssim 0.6$ this
ratio becomes the same for different bosons, whereas at large
$x_1\to 1$ the relative contribution of the longitudinally polarized
photon dominates in corresponding ratios for other bosons.
\begin{figure}[!h]
\begin{minipage}{0.49\textwidth}
 \centerline{\includegraphics[width=1.0\textwidth]{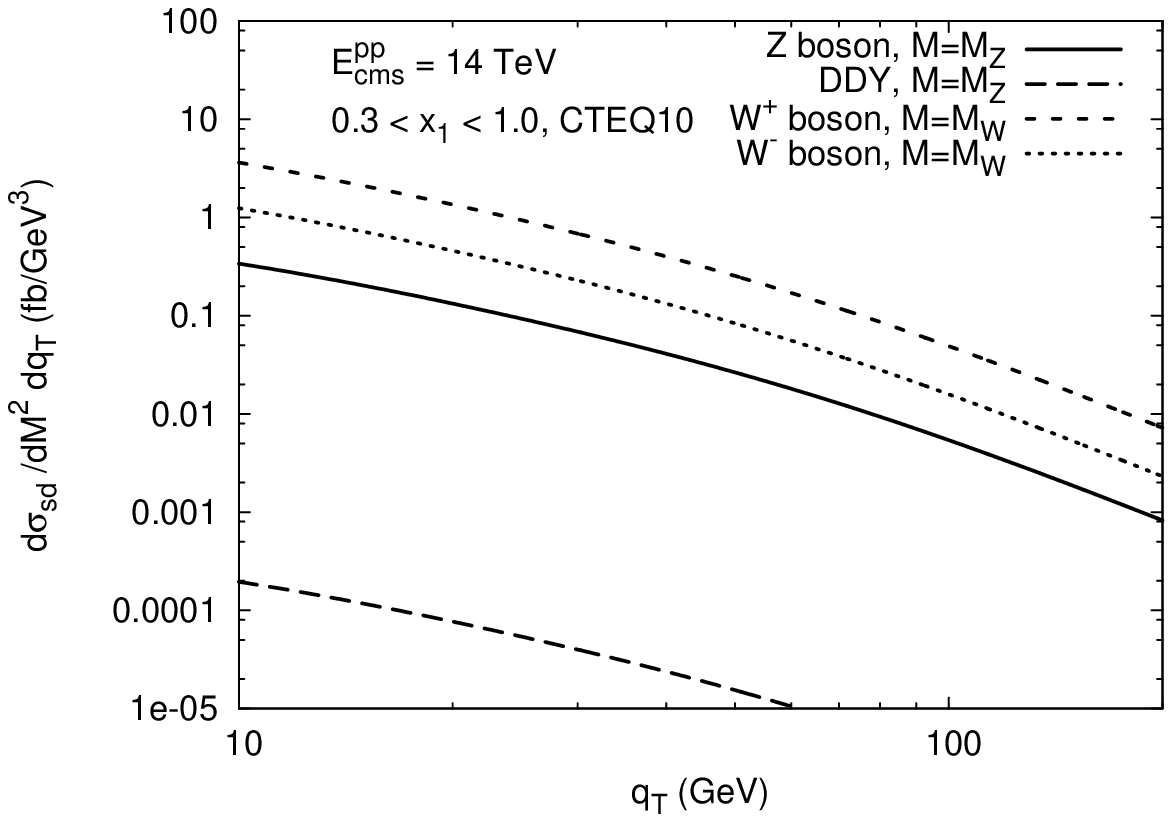}}
\end{minipage}
\hspace{0.5cm}
\begin{minipage}{0.46\textwidth}
 \centerline{\includegraphics[width=1.0\textwidth]{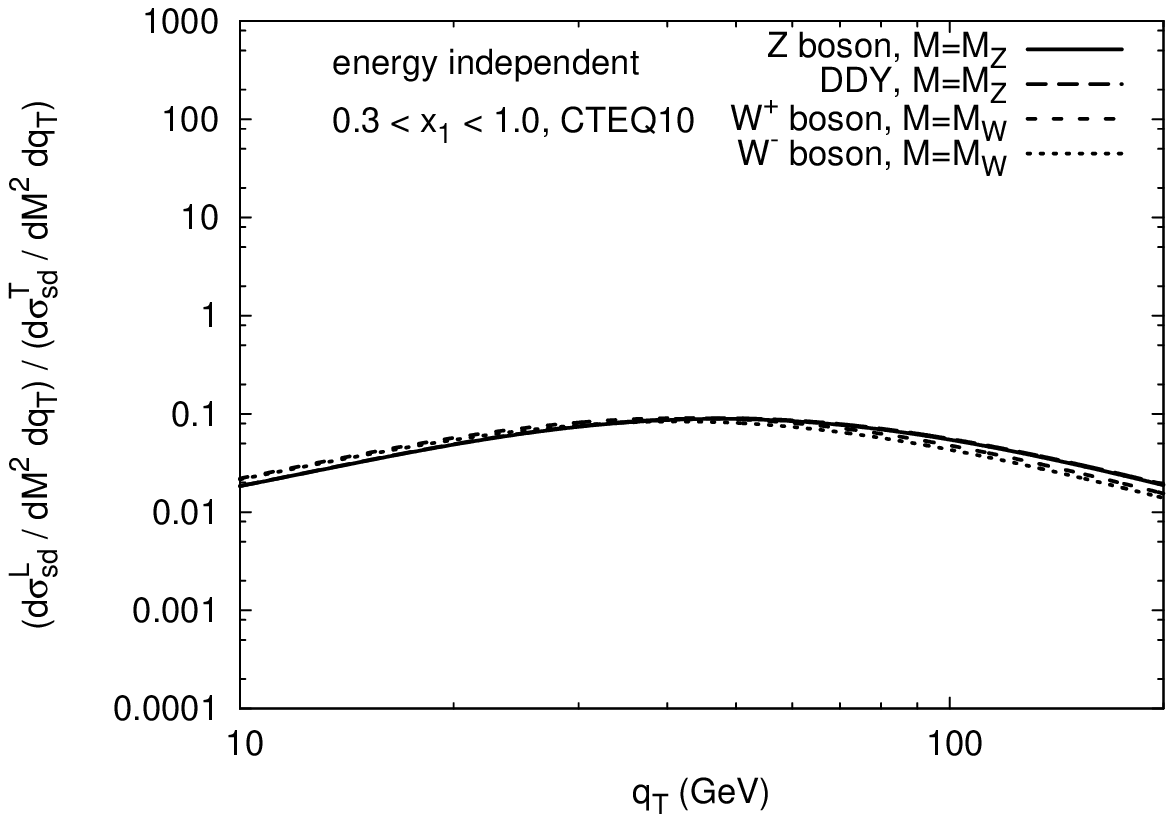}}
\end{minipage}
   \caption{
\small The di-lepton transverse momentum $q_{\perp}$ distribution of
the doubly-differential diffractive cross section at the LHC energy
$\sqrt{s}=14$ TeV at fixed di-lepton invariant mass is shown in
the left panel. The longitudinal-to-transverse gauge bosons
polarisations ratio as a function of the di-lepton $q_{\perp}$ is
shown in the right panel. In both panels, the invariant mass is
fixed as $M=M_Z$ in the $Z^0,\gamma^*$ production case and as
$M=M_W$ in the $W^{\pm}$ production case. CTEQ10 PDF parametrization
\cite{CTEQ10} is used here.}
 \label{fig:dqt}
\end{figure}

From the phenomenological point of view, the distribution of the
forward diffractive cross section in the di-lepton transverse
momentum $q_{\perp}$ could also be of major
importance\footnote{Authors are indebted to Torbj\"orn Sj\"ostrand
for pointing out this point.}. In Fig.~\ref{fig:dqt} (left
panel) we show the di-lepton transverse momentum $q_{\perp}$
distribution of the doubly-differential diffractive cross section at
the LHC energy $\sqrt{s}=14$ TeV at the di-lepton invariant
mass, fixed at a corresponding resonance value -- the $Z$ or $W$
mass. The shapes turned out to be smooth and the same for different
gauge bosons, and are different mostly in normalisation. In
Fig.~\ref{fig:dqt} (right panel) we show the $q_{\perp}$ dependence
of the $\sigma^L/\sigma^T$ ratio in the resonances. We notice that
the ratio does not strongly vary for different bosons. It is peaked
at about the half of the resonance mass, and uniformly decreases to
smaller/larger $q_{\perp}$ values.
\begin{figure*} [!h]
\begin{minipage}{0.49\textwidth}
 \centerline{\includegraphics[width=1.0\textwidth]{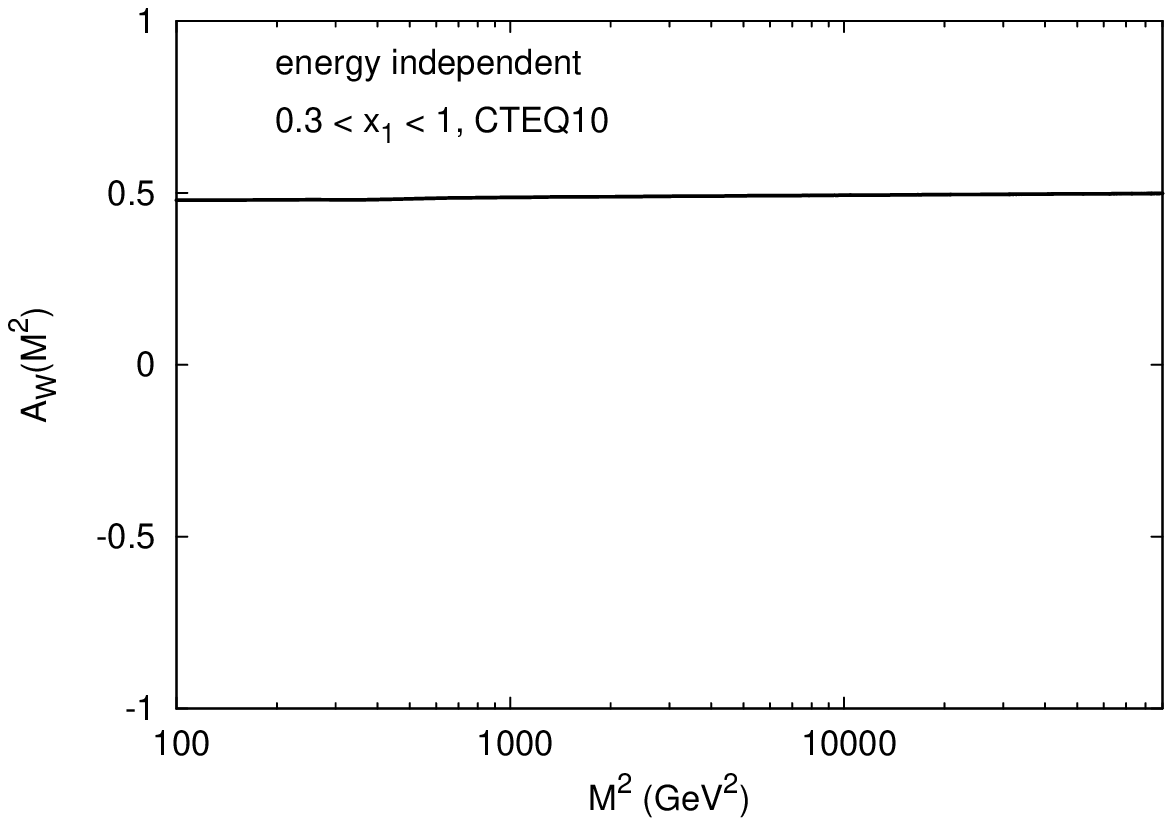}}
\end{minipage}
\hspace{0.2cm}
\begin{minipage}{0.48\textwidth}
 \centerline{\includegraphics[width=1.0\textwidth]{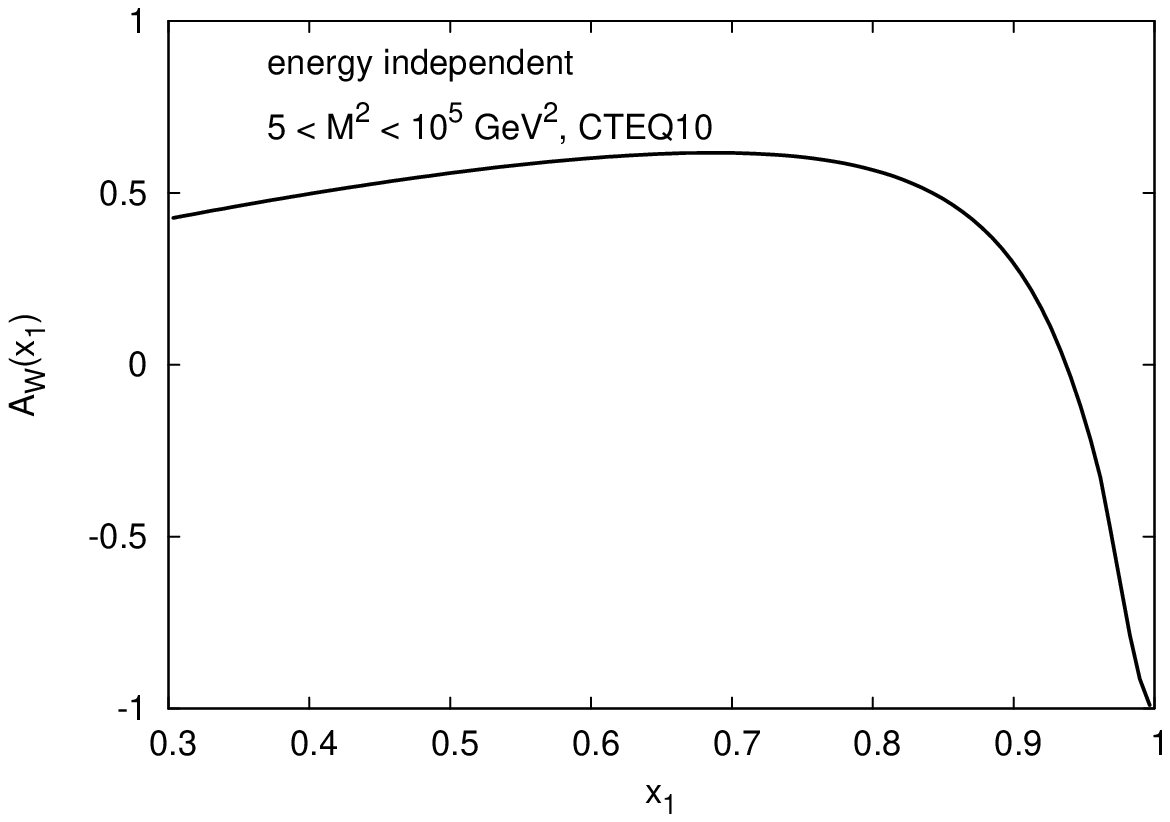}}
\end{minipage}
   \caption{
\small Charge asymmetry in the single diffractive $W^+$ and $W^-$
cross sections as a function of $M^2$, at fixed $x_1=0.5$ (left
panel), and $x_1$, at fixed $M^2=M_W^2$ (right panel). Solid lines
correspond to the LHC energy $\sqrt{s}=14$ TeV, dished lines -- to
the RHIC energy $\sqrt{s}=500$ GeV.}
 \label{fig:AW}
\end{figure*}

As one of the important observables, sensitive to the difference
between $u$- and $d$-quark PDFs, the $W^{\pm}$ charge asymmetry
$A_W$ is shown in Fig.~\ref{fig:AW} differentially as a function of
the di-lepton invariant mass squared $M^2$ and integrated over
$0.3<x_1<1.0$ interval (left panel)
\begin{eqnarray}
A_W(M^2)=\frac{d\sigma_{sd}^{W^+}/dM^2-d\sigma_{sd}^{W^-}/dM^2}
{d\sigma_{sd}^{W^+}/dM^2+d\sigma_{sd}^{W^-}/dM^2}\,,
\end{eqnarray}
and as a function of the boson momentum fraction $x_1$ and
integrated over $5<M^2<10^5$ GeV$^2$ interval (right panel)
\begin{eqnarray}
A_W(x_1)=\frac{d\sigma_{sd}^{W^+}/dx_1-d\sigma_{sd}^{W^-}/dx_1}
{d\sigma_{sd}^{W^+}/dx_1+d\sigma_{sd}^{W^-}/dx_1}\,.
\end{eqnarray}
The ratio turns out to be independent on both the hard scale $M^2$
and the c.m. energy $\sqrt{s}$. One concludes that, due to
different $x$-shapes of valence $u,\,d$ quark PDFs, at smaller
$x_1\lesssim 0.9$ the diffractive $W^+$ bosons' rate dominates over
$W^-$ one. However, at large $x_1\to 1$ the $W^-$ boson cross
section becomes increasingly important and strongly dominates over
the $W^+$ one.
\begin{figure*} [!h]
\begin{minipage}{0.49\textwidth}
 \centerline{\includegraphics[width=1.0\textwidth]{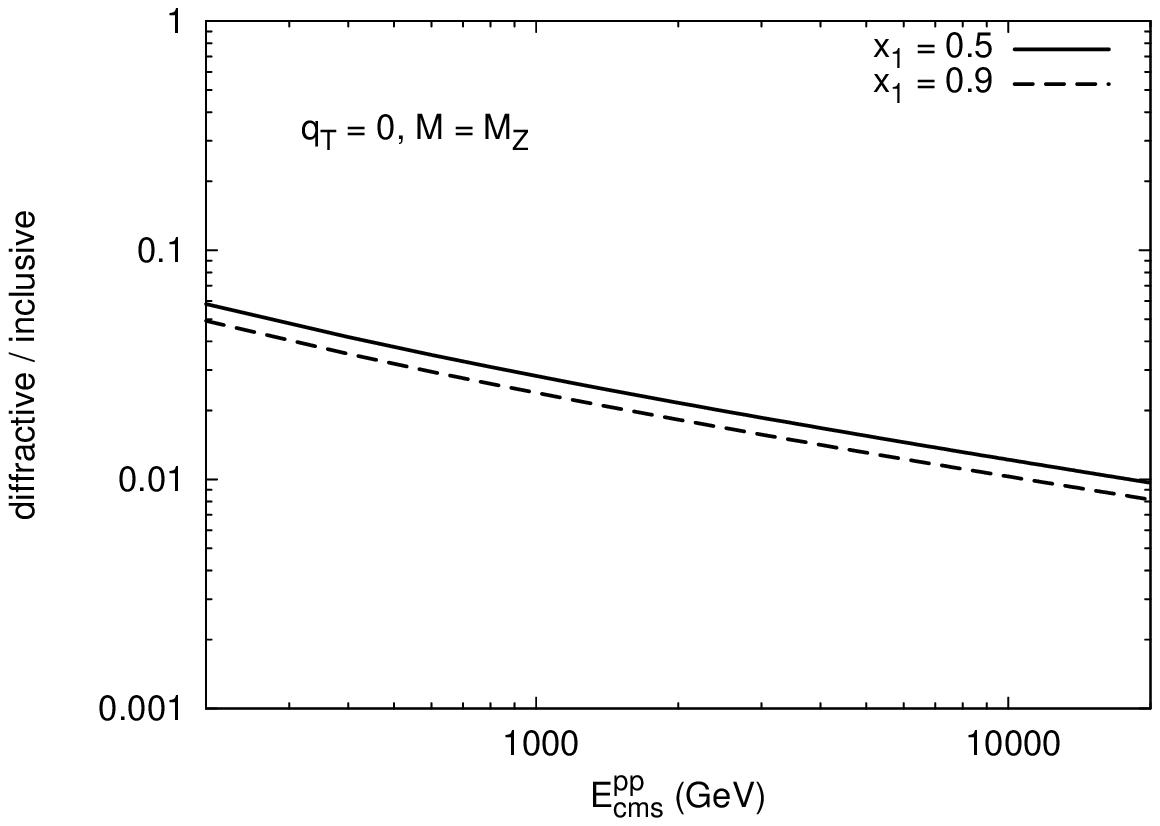}}
\end{minipage}
\hspace{0.2cm}
\begin{minipage}{0.48\textwidth}
 \centerline{\includegraphics[width=1.0\textwidth]{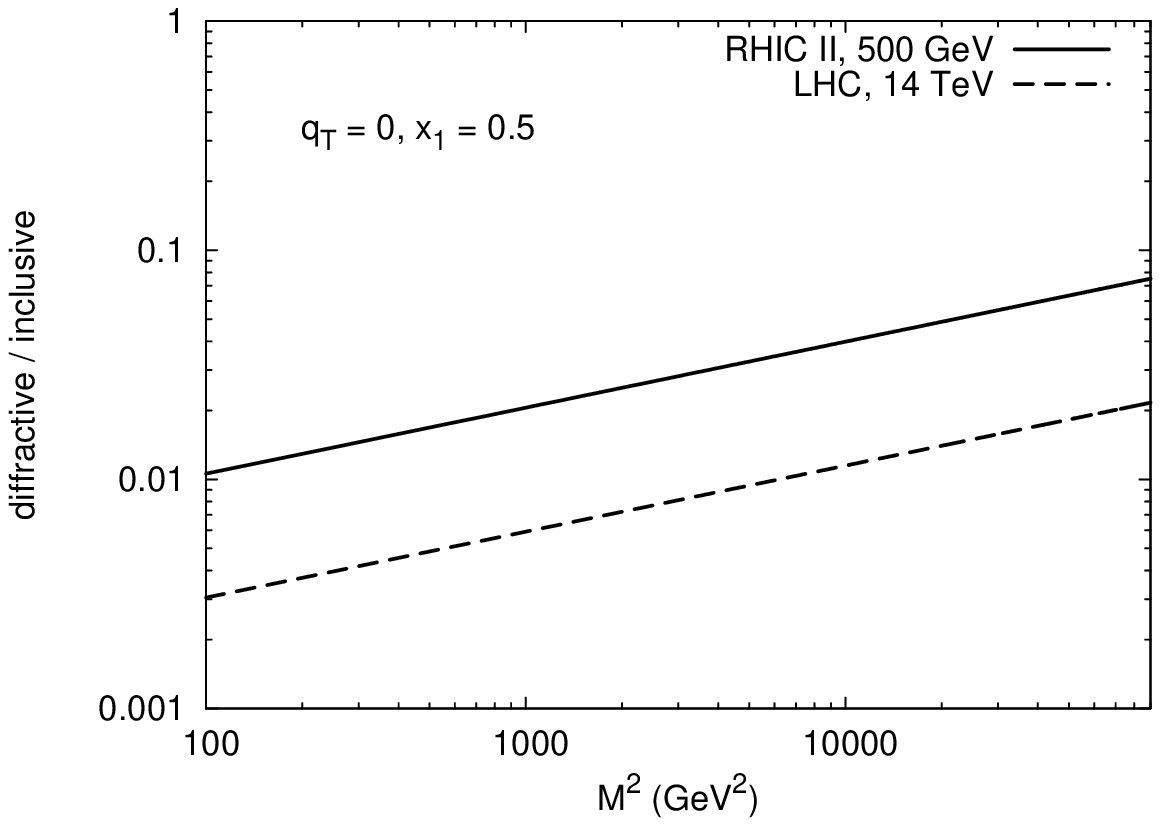}}
\end{minipage}
   \caption{
\small The diffractive-to-inclusive ratio of the gauge bosons
production cross sections in $pp$ collisions derived in
Eq.~(\ref{ratDDincl}) as a function of the c.m. energy
$\sqrt{s}$ (left panel) and the di-lepton invariant mass $M^2$
(right panel). It does not depend on the type of the gauge boson and
quark PDFs.}
 \label{fig:ratio1}
\end{figure*}

An important feature of the diffractive-to-inclusive Abelian
radiation cross sections ratio
\begin{eqnarray}\label{Rrat}
R(M^2,x_1)=\frac{d\sigma_{sd}/dx_1dM^2} {d\sigma_{incl}/dx_1dM^2}\,,
\end{eqnarray}
which makes these predictions different from ones obtained in
traditional diffractive QCD factorisation-based approaches (see e.g.
Refs.~\cite{Szczurek,Beatriz}), is their unusual energy and scale
dependence demonstrated in Fig.~\ref{fig:ratio1}. Notice that we
stick to the case of small boson transverse momenta, $q_{\perp}\ll
M$, where the main bulk of diffractive signals comes from. The
analytic formula for this ratio was derived above and is shown by
Eq.~(\ref{ratDDincl}), which demonstrates that the ratio is
independent of the type of the gauge boson, its polarisation, or
quark PDFs. In this respect, it is the most convenient and model
independent observable, which is sensitive only to the structure of
the universal elastic dipole amplitude (or the dipole cross
section), and can be used as an important probe for the QCD
diffractive mechanism for forward diffractive reactions, essentially
driven by the soft interaction dynamics. We see from
Fig.~\ref{fig:ratio1} that the $\sigma_{sd}/\sigma_{incl}$ ratio
decreases with energy, but increases with the hard scale, thus
behaves opposite to what is expected in the diffractive
factorisation-based approaches. Therefore, measurements of the
single diffractive gauge boson production cross section, at least,
at two different energies would provide important information about
the interplay between soft and hard interactions in QCD, and its
role in formation of diffractive excitations and color screening
effects.
\begin{figure*}[!h]
\begin{minipage}{0.49\textwidth}
 \centerline{\includegraphics[width=1.0\textwidth]{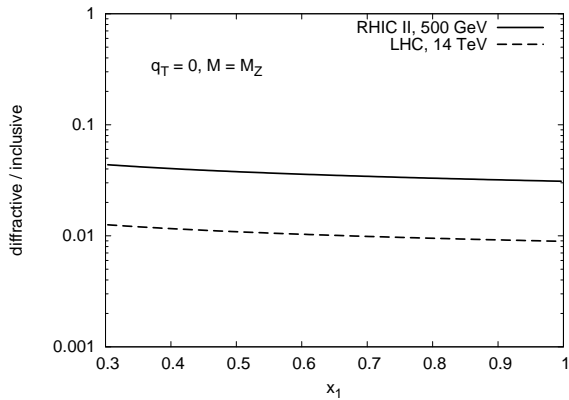}}
\end{minipage}
   \caption{
\small The diffractive-to-inclusive ratio as function of the boson
fractional momentum $x_1$. }
 \label{fig:ratio2}
\end{figure*}

Finally, in Fig.~\ref{fig:ratio2} we present the
diffractive-to-inclusive cross sections ratio as a function of the
boson fractional momentum $x_1$ at RHIC ($\sqrt{s}=500$ GeV) and at
LHC ($\sqrt{s}=14$ TeV) energies (left panel). In the right panel,
we compare this ratio calculated at the Tevatron energy
($\sqrt{s}=1.96$ TeV) with the recent measurements of diffractive
$W$ and $Z$ production performed by the CDF collaboration
\cite{CDF-WZ}. The data show the $x_1$-integrated ratio of
diffractive to inclusive cross sections. Due to the weak
$x_1$-dependence of this ratio (left panel) with a good accuracy the
integrated values are numerically close to the ratio of differential
cross sections given by Eq.~(\ref{Rrat}). We see that the results of
our calculations with Eq.~(\ref{ratDDincl}) at $x_1=0,5$ agree well
with the data. This agreement is another confirmation of correctness
of the absorption effects included into the parametrization of the
dipole cross section (\ref{KST}).

\section{The results meet data}

\subsection{Link to the Regge phenomenology and  data}

\begin{figure*}[!h]
\begin{minipage}{0.8\textwidth}
 \centerline{\includegraphics[width=1.0\textwidth]{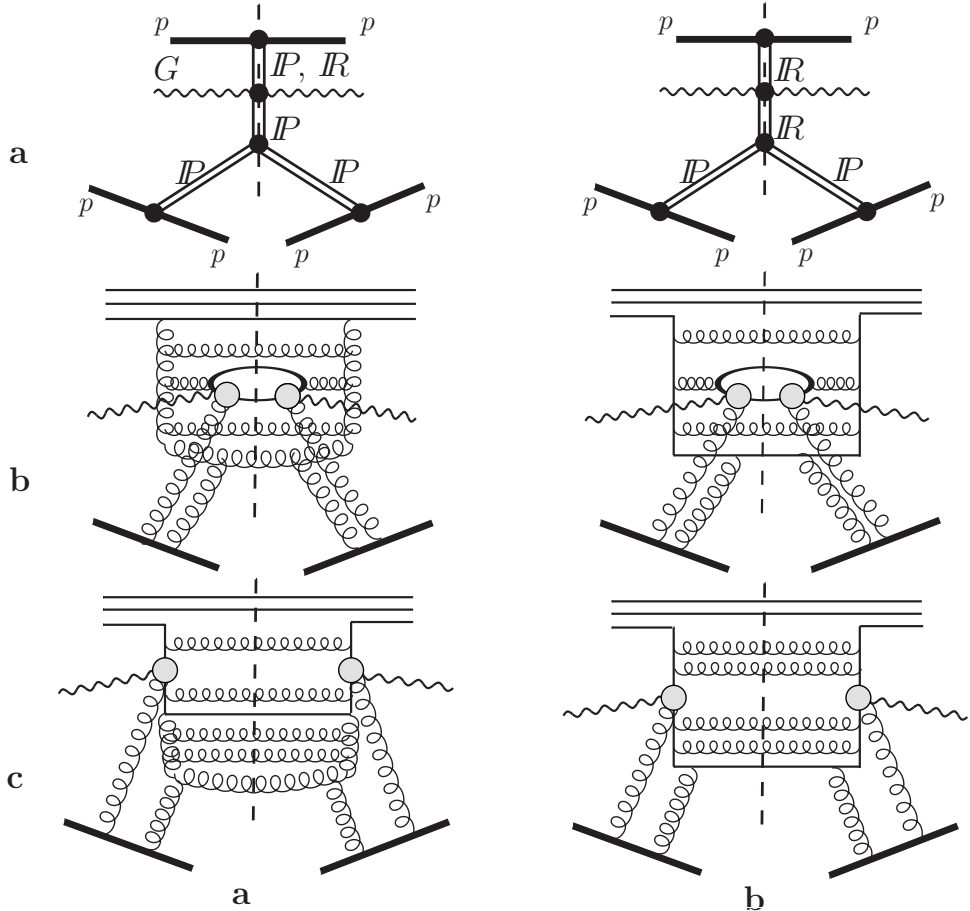}}
\end{minipage}
   \caption{
\small {\it The upper row:} the triple-Regge graphs for the process
$pp\to Xp$, where the diffractively produced state $X$ contains a
gauge boson. Examples of Feynman graphs corresponding to diffractive
excitation of a large invariant mass, going along with radiation of
a gauge boson are displayed in {\it the 2d and 3rd rows}. Curly and
waving lines show gluons and the radiated gauge boson. The dashed
line indicates the unitarity cut.}
 \label{fig:3-regge}
\end{figure*}

\begin{figure*}[!h]
\begin{minipage}{1.0\textwidth}
 \centerline{\includegraphics[width=0.7\textwidth]{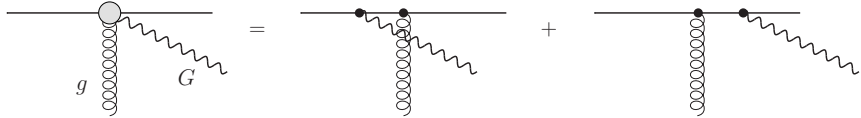}}
\end{minipage}
   \caption{The effective amplitude of gauge boson
   radiation by a projectile quark.}
 \label{fig:vertex}
\end{figure*}

The process $pp\to Xp$ at large Feynman $x_F\to1$ of the recoil
proton, or small \beq \xi=1-x_F=\frac{M_X^2}{s}\ll1, \label{xi} \eeq
 is described by triple Regge graphs $\Pom\Pom\Pom$ and $\Pom\Pom\Reg$
 depicted in Fig.~\ref{fig:3-regge}, (aa)
 and (ab) respectively, were we also included radiation of a gauge boson.
Examples of Feynman graphs corresponding to the above triple-Regge
terms,  are shown in the second and third rows in
Fig.~\ref{fig:3-regge}. The graphs (ba) and (ca) illustrate the
triple-Pomeron term in the diffraction cross section, \beq
\frac{d\sigma_{diff}^{\Pom\Pom\Pom}}{d\xi dt} \propto
\xi^{-\alpha_{\Pom}(0)-2\alpha^\prime_{\Pom}(t)}, \label{PPP} \eeq
with the gauge boson radiated by either a sea, (ba), or a valence
quark, (ca). The effective radiation amplitude $q+g\to q+G$ is
depicted by open circles and is defined in Fig.~\ref{fig:vertex}.
These Feynman graph interpret the triple-Pomeron term as a
diffractive excitation of the incoming proton due to radiation of
gluons with small fractional momentum. The proton can also
dissociate via diffractive excitation of its valence quark skeleton,
as is illustrated in Fig.~\ref{fig:3-regge} (bb) and (cb). The
corresponding term in the diffraction cross section reads, \beq
\frac{d\sigma_{diff}^{\Pom\Pom\Reg}}{d\xi dt} \propto
\xi^{\alpha_{\Reg}(0)-\alpha_{\Pom}(0)-2\alpha^\prime_{\Pom}(t)},
\label{PPR} \eeq Again, the gauge boson can be radiated either by a
sea quark, (bb), or by the valence quark, (cb).

It is worth emphasizing that the quark radiating the gauge boson
cannot be a spectator, but must participate in the interaction. This
is a straightforward consequence of the Good-Walker mechanism of
diffraction \cite{GW}. As was discussed above, the contribution of a
given projectile Fock state to the diffraction amplitude is given by
the difference of elastic amplitudes for the Fock states including
or excluding the gauge boson, \beq \Im f^{(n)}_{diff}=\Im
f^{(n+G)}_{el}-\Im f^{(n)}_{el}, \label{f-diff} \eeq where $n$ is
the total number of partons in the Fock state; $f^{(n+G)}_{el}$ and
$f^{(n)}_{el}$ are the elastic scattering amplitudes for the whole
$n$-parton ensemble, which either contains the gauge boson or does
not, respectively. Although the gauge boson does not participate in
the interaction, the impact parameter of the quark radiating the
boson gets shifted, and this is the only reason why the difference
Eq.~(\ref{f-diff}) is not zero. This also conveys that this quark
must interact in order to retain the diffractive amplitude nonzero
\cite{KPST06}. For this reason in the graphs depicted in
Fig.~\ref{fig:3-regge}  the quark radiating $G$ always takes part in
the interaction with the target.

Notice that there is no one-to-one correspondence between
diffraction in QCD and the triple-Regge phenomenology. In
particular, there is no triple-Pomeron vertex localized in rapidity.
The colorless "Pomeron" contains at least two $t$-channel gluons,
which can couple to any pair of projectile partons. For instance in
diffractive gluon radiation, which is the lowest order term in the
triple-Pomeron graph, one of the $t$-channel gluons can couple to
the radiated gluon, while another one couples to another parton at
any rapidity, e.g. to a valence quark (see Fig.~3 in
\cite{KST-par}). Apparently, such a contribution cannot be
associated literally with either of the Regge graphs in
Fig.~\ref{fig:3-regge}. Nevertheless, this does not affect much the
$x_F$- and energy dependencies provided by the triple-Regge graphs,
because the gluon has spin one.

It is also worth mentioning  that in Fig.~\ref{fig:3-regge} we
presented only the lowest order  graphs with two gluon exchange. The
spectator partons in a multi-parton Fock component also can interact
and contribute to the elastic amplitude of the whole parton
ensemble. This gives rise to higher order terms, not shown
explicitly in Fig.~\ref{fig:3-regge}. They contribute to the
diffractive amplitude Eq.~(\ref{f-diff}) as a factor, which we
define as the gap survival amplitude.

\subsection{Gap survival amplitude}

The amplitude of survival of a large rapidity gap is controlled by
the largest dipoles in the projectile hadron.  This was included in
our evaluation of the diffractive amplitude Eq.~(\ref{Delta}). Soft
gluons in the light-cone wave function of the proton should also be
considered as spectator partons, and the large (compared with
$1/M_G$) distance $R_{ij}$ in Eq.~(\ref{lim}) in this case is the
quark-gluon separation. In fact, our calculations do include such
configurations. Indeed, data on diffraction show that diffractive
gluon radiation is quite weak (well known smallness of the
triple-Pomeron coupling), and this can be explained assuming that
gluons in the proton are located within small ``spots'' around the
valence quarks  with radius $r_0\sim 0.3\fm$
\cite{KST-par,shuryak,shuryak-zahed,spots}. Therefore, the large
distance between one valence quark and a satellite-gluon of the
other quark is approximately equal (with 10\% accuracy) to the
quark-quark separation. Since a valence quark together with
co-moving gluons is a color triplet, in our calculations the
interaction amplitude of such an effective (``constituent'') quark
with the target is a coherent sum of the quark-target and
gluon-target interaction amplitudes.

In addition to the soft gluons, which are present in the proton
light-cone wave function at a soft scale, production of a heavy
gauge boson certainly lead to an additional intensive hard gluon
radiation. In other words, there might be many more spectator gluons
in the quark which radiates the gauge boson. The transverse
separation of those gluons is controlled by the DGLAP evolution.
One can replace a bunch of gluons by dipoles \cite{mueller} which
transverse size $r_d$ varies from $1/M_G$ up to $r_0$, and is
distributed as $dr_d/r_d$ \cite{nz94}. Therefore the mean dipole
size squared, \beq \la r_d^2\ra = \frac{r_0^2}{\ln(r_0^2M_G^2)},
\label{mean-dip} \eeq is about $\la r_d^2\ra\approx 0.01\fm^2$, i.e.
quite small. The cross section of such a dipole on a proton is also
small, $\sigma_d=C(x)\,\la r_d\ra^2$, where according to
Eq.~(\ref{fel}) factor $C(x)=\sigma_0/R_0^2(x)$ rises with energy.
Fixing $x=M_G^2/s$ and using the parameters fitted  in \cite{GBWdip}
to DIS data from HERA we get at the Tevatron collider energy
$\sigma_d\approx 0.9\mb$.

Presence of each such a dipole in the projectile light-cone wave
function brings an extra suppression factor to the survival
amplitude of a large rapidity gap, \beq S_d(s)=1-\Im f_d(b,r_d)
\label{S_d} \eeq We aimed here at a demonstration that the second
term in (\ref{S_d})  is negligibly small, so we rely on its
simplified form (see more involved calculations in \cite{dis}), \beq
 \Im f_d(b,r_d)\approx \frac{\sigma_d}{4\pi B_d}\,e^{-b^2/2B_d},
 \label{f_d}
 \eeq
where $B_d$ is the dipole-nucleons elastic slope, which was measured
at $B_d\approx6\GeV^{-2}$ in diffractive electro-production of
$\rho$ mesons at HERA \cite{rho}.

We evaluate the absorptive correction (\ref{f_d}) at the mean impact
parameter $\la b^2\ra=2B_d$ and for the Tevatron energy
$\sqrt{s}=2\TeV$ arrive at the negligibly small value $\Im
f_d(0,r_d)\approx 0.01$.

However, the number of such dipole rises with hardness of the
process,and may substantially enhance the magnitude of the
absorptive corrections. The gap survival amplitude for $n_d$
projectile dipoles reads, \beq S^{(n_d)}_d=\bigl[1-\Im
f_d(b,r_d)\bigr]^{n_d}. \label{S_n} \eeq

The mean number of dipoles can be estimated in in the
double-leading-log approximation to the DGLAP evolution formulated
in impact parameters \cite{nz94}, the mean number of such dipoles is
given by \beq \la n_d\ra = \sqrt{\frac{12}{\beta_0}
\ln\left(\frac{1}{\alpha_s(M_G^2)}\right) \ln\left((1-x_F){s\over
s_0}\right) }. \label{mean-n} \eeq Here the values of Bjorken $x$ of
the radiated gluons is restricted by the invariant mass of the
diffractive excitation, $x>s_0/M_X^2=s_0/(1-x_F)s$. For the
kinematics of experiments at the Tevatron collider (see next
section), $1-x_F<0.1$, $\sqrt{s}=2\TeV$, the number of radiated
dipoles is not large, $\la n_d\ra\lesssim6$. We conclude that the
absorptive corrections Eq.~(\ref{S_n}) to the gap survival amplitude
are rather weak, less than $5\%$, i.e. about $10\%$ in the survival
probability. This correction is certainly small compared to other
theoretical uncertainties of our calculations. Notice that a similar
correction due to radiation of soft gluons was found in \cite{dis}
for the gap survival probability in leading neutron production in
DIS.

\subsection{Comparison with data}

\begin{figure*}[!h]
\begin{minipage}{0.48\textwidth}
 \centerline{\includegraphics[width=1.0\textwidth]{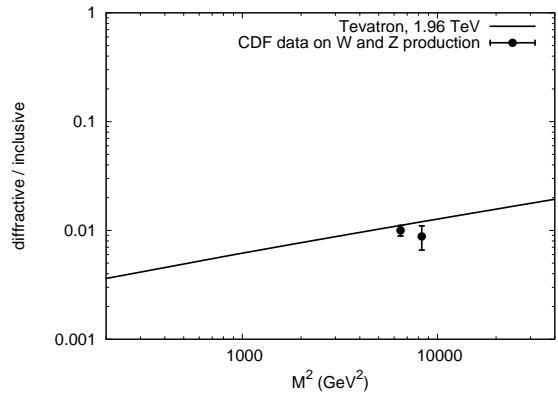}}
\end{minipage}
   \caption{
\small The diffractive-to-inclusive ratio as function of the
invariant mass squared of the produced dilepton. The CDF data for
$W$ and $Z$ production were taken at the Tevatron energy
($\sqrt{s}=1.96$ TeV). The first CDF data point corresponds to the
$W$ production, $M^2=M_W^2$, the second -- to the $Z$ production,
$M^2=M_Z^2$.}
 \label{fig:ratio3}
\end{figure*}

Thus, our calculations effectively cover the gluon radiation, so the
triple-Pomeron term  is included. This is important because this
term dominates the diffractive cross section \cite{3R}. So we can
compare with available data from the CDF experiment \cite{CDF-WZ} on
$W$ and $Z$ diffractive production depicted in
Fig.~\ref{fig:ratio3}.

However, in order to compare our results with CDF data, we have to
introduce in our calculations the proper experimental cuts,
namely, $0.03<\xi\equiv1-x_F<0.1$ \cite{CDF-WZ}. Since our diffractive cross
section formulae are differential in $M^2$, not in $M_X^2$, and
experimental cuts on $y$-rapidity distribution of a produced gauge
boson are unavailable at the moment, a direct implementation of the
$\xi$ cuts into our formalism cannot be performed immediately.

As a way out of this problem, at small $\xi\to 0$ one can instead
write the single diffractive cross section in the phenomenological
triple-Regge form \cite{3R},
 \beqn
- \frac{d\sigma_{sd}^{pp}}
{d\xi\,dp_T^2} =
\sqrt{\frac{s_1}{s}}\,
\frac{G_{\Pom\Pom\Reg}(0)}
{\xi^{3/2}}\,
e^{-B_{\Pom\Pom\Reg}p_T^2} +
\frac{G_{3\Pom}(0)}
{\xi}\,
e^{-B^{pp}_{3\Pom}p_T^2}\,,
\label{140}
 \eeqn
where $s_1=1\,\GeV^2$; $B_{\Pom\Pom i} = R^2_{\Pom\Pom i} -
2\alpha^\prime_\Pom\,\ln\xi$; $i=\Pom,\Reg$;  and
$\alpha^\prime_\Pom\approx 0.25\GeV^{-2}$ is the slopes of the
Pomeron trajectory. Then an effect of the experimental cuts on $\xi$
in the phenomenological cross section (\ref{140}) and in our
diffractive cross section calculated above (\ref{eik-tot}) should
roughly be the same.

Since the data show no substantial rise of the diffractive cross
section with energy \cite{dino-sd,peter}, which is apparently caused
by strong absorptive corrections, we incorporate this fact fixing
the effective Pomeron intercept at $\alpha_\Pom(0)=1$. This also
allows us to use the results of the comprehensive triple-Regge
analysis of data performed in Ref.~\cite{3R}, which led to the
following values of the parameters: $G_{3\Pom}(0)=
G_{\Pom\Pom\Reg}(0)=3.2\,\mb/\GeV^2$; $R^2_{3\Pom}=4.2\,\GeV^{-2}$;
$R^2_{\Pom\Pom\Reg}=1.7\,\GeV^{-2}$.

Now we are in a position to evaluate the suppression factor $\delta$
caused by experimental cut on $\xi$, by taking the ratio
 \beq \delta=\frac{\int dp_T^2\int_{0.03}^{0.1}
d\xi\,d\sigma/dp_T^2 d\xi} {\int dp_T^2\int_{\xi_{min}}^{\xi_{max}}
d\xi\,d\sigma/dp_T^2 d\xi} \label{delta}
 \eeq Here
$\xi_{min}=M_{X,min}^2/s$, where $M_{X,min}\simeq M_Z$ is the
minimal produced diffractive mass containing a heavy gauge boson.
The value of $\delta$ in Ref.~(\ref{delta}) is essentially
determined by the experimental cuts on $\xi$ and is not sensitive to
the upper limit $\xi_{max}$ in denominator, so we fix it at a
realistic value\footnote{The estimate for $\xi_{max} \sim 0.3$
corresponds to the limiting case when one of the constituent quarks
in a target (anti)proton looses almost all its energy into a hard
radiation of a gluon in the $t$-channel. The second and all
subsequent $t$-channel gluon exchanges collectively screen the color
charge taken away from the target by the first gluon and transfer
much smaller fraction of initial target momentum to projectile
quarks as has been recently advocated in
Refs.~\cite{screen1,screen2}.}: $\xi_{max}\sim0.3$. Then
Eq.~(\ref{delta}) leads to $\delta\simeq 0.2$, the factor reducing
the diffractive gauge bosons production cross section calculated
above. Our result plotted in Fig.~\ref{fig:ratio3} demonstrates a
good agreement with the CDF data on single diffractive $W$ and $Z$
production \cite{CDF-WZ}.

\section{Conclusions}

The diffractive radiation of Abelian fields, $\gamma$, $Z^0$,
$W^{\pm}$, expose unusual features, which make it very different
from diffraction in DIS, and lead to a dramatic breakdown of QCD
factorisation in diffraction.

The first, rather obvious source for violation of diffractive
factorisation is related to absorptive corrections (called sometimes
survival probability of large rapidity gaps). The absorptive
corrections affect differently the diagonal and off-diagonal terms
in the hadronic current \cite{PCAC},  leading to an unavoidable
breakdown of QCD factorisation in processes with off-diagonal
contributions only. Namely, the absorptive corrections suppress the
off-diagonal diffraction much stronger than the diagonal channels.
In the diffractive Abelian radiation in hadron-hadron collisions a
new state, i.e. the gauge boson decaying into the heavy lepton pair,
is produced, hence, the whole process is of entirely off-diagonal
nature, whereas in the diffractive DIS contains both diagonal and
off-diagonal contributions \cite{KPSdiff}. This is the first reason
why QCD factorisation is broken in the diffractive gauge bosons
production processes.

The second, more sophisticated reason to contradict diffractive
factorisation is specific for Abelian radiation, namely, a quark
cannot radiate in the forward direction (zero momentum transfer),
where diffractive cross sections usually have a maximum. Forward
diffraction becomes possible due to intrinsic transverse motion of
quarks inside the proton.

Third, the mechanism of Abelian radiation in the forward direction
in $pp$ collisions is related to participation of the spectator
partons in the proton. Namely, the perturbative QCD interaction of a
projectile quark is responsible for the hard process of a heavy
boson radiation, while a soft interaction with the projectile
spectator partons provides color neutralization
\cite{screen1,screen2}, which is required for a diffractive (Pomeron
exchange) process. Such an interplay of hard and soft dynamics is
also specific for the process under consideration, which makes it
different from the diffractive DIS, dominated exclusively by soft
interactions, and which also results in breakdown of diffractive
factorisation.

The diffractive (Ingelman-Schlein) QCD factorisation breaking
manifests itself in specific features of diffractive observables
like a significant damping of the single diffractive gauge bosons
production cross section at high $\sqrt{s}$ compared to the
inclusive production case. This is rather unusual, since a
diffractive cross section, which is proportional to the dipole cross
section squared, could be expected to rise with energy steeper than
the total inclusive cross section, like it occurs in the diffractive
DIS process. At the same time, the ratio of the single diffractive
to inclusive production cross sections rises with the hard scale,
$M^2$. This is also in variance with diffraction in DIS associated
with the soft interactions.

In this paper, we have presented the differential distributions (in
transverse momentum, invariant mass and longitudinal momentum
fraction) of the diffractive $\gamma^*,\,Z^0$ and $W^{\pm}$ bosons
production at RHIC (500 GeV) and LHC (14 TeV) energies, as well
as the ratio of the boson longitudinal to transverse polarisation
contributions. We have also calculated the charge $W^{\pm}$
asymmetry, relevant for upcoming measurements at the LHC. The ratio
diffractive to inclusive gauge bosons production cross sections does
not depend on a particular type of the gauge boson, its polarisation
state and quark PDFs, and depends only on properties of the
universal dipole cross section and sensitive to the saturation scale
at small $x$. Finally, our prediction for this ratio is numerically
consistent with the one measured for diffractive $W$ and $Z$
production at the Tevatron.

The theoretical uncertainties of our calculations come mainly from
poorly known quark PDFs (or the proton structure function $F_2(x_q)$
in diffractive DY) at large quark fractions $x_q\to 1$. This issue
has been discussed in our previous analysis of the DDY process in
Ref.~\cite{our-DDY}, where a strong sensitivity of the $x_1$
dependence to a particular $F_2$ parametrization has been pointed
out. The same situation extends to the more general case of
diffractive Abelian radiation considered in this paper. This tells
us again that measurements of forward diffractive gauge bosons
production would be extremely important or even crucial for settling
further more stringent constraints on the quark content of the
proton.\\

{\bf Acknowledgments}

Useful discussions and helpful correspondence with Gunnar Ingelman,
Valery Khoze, Yuri Kovchegov, Eugene Levin, Amir Rezaeian, Christophe Royon,
Torbj\"orn Sj\"ostrand and Antoni Szczurek are gratefully
acknowledged. This study was partially supported by Fondecyt (Chile)
grant 1090291, and by Conicyt-DFG grant No. 084-2009.


\end{document}